\documentclass[authoryear,table]{article}

\usepackage{arxiv}
\usepackage{amssymb}
\usepackage{soul}
\usepackage{tikz}
\usepackage{pgfplots} 
\pgfplotsset{compat=1.16}
\usepackage[most]{tcolorbox}

\usepackage{subfigure}
\usepackage{natbib}
\usepackage{titlecaps}

\usepackage[hang]{footmisc}
\fancypagestyle{plain}{%
  \fancyhf{}
  \fancyhead[R]{\small {\it \jname}, x(x):\thepage--\pageref{LastPage} (\myyear\today)}
  \fancyfoot[R]{\small\bf\thepage }
  \fancyfoot[L]{\fottitle}
  }
\renewcommand{\today}{}

\usepackage{url}
\usepackage{xcolor} 
\definecolor{azure}{rgb}{0.0, 0.5, 1.0}
\definecolor{darkgreen}{rgb}{0.0, 0.5, 0.0}
\definecolor{amaranth}{rgb}{0.9, 0.17, 0.31}
\definecolor{cadetgrey}{rgb}{0.57, 0.64, 0.69}
\definecolor{aureolin}{rgb}{0.99, 0.93, 0.0}

\usepackage{listings}

\usepackage{url}

\usepackage{csquotes}

\usepackage{booktabs}

\usepackage{hyperref}

\hypersetup{
    colorlinks=true,
    citecolor=blue,
}
\newcommand{\cpp}[0]{C\texttt{++}}
\newcommand{\tikzxmark}{%
\tikz[scale=0.23] {
    \draw[line width=0.7,line cap=round] (0,0) to [bend left=6] (1,1);
    \draw[line width=0.7,line cap=round] (0.2,0.95) to [bend right=3] (0.8,0.05);
}}
\newcommand{\newpara}
    {
    \vskip 0.1cm
    }

\newcommand{\change}[1]{\textcolor{black}{#1}}

\begin{document}

\title{LLM Benchmarking with LLaMA2: Evaluating Code Development Performance Across Multiple Programming Languages}


\author{Patrick Diehl, Maxim Moraru\\Los Alamos National Laboratory, Los Alamos, NM, 87544, USA \AND Nojoud Nader, Steven R. Brandt \\
Louisiana State University, Baton Rouge, LA, 70803, USA 
}
\maketitle

\begin{abstract}
The rapid evolution of large language models (LLMs) has opened new possibilities for automating various tasks in software development. This paper evaluates the capabilities of the Llama 2-70B model in automating these tasks for scientific applications written in commonly used programming languages. Using representative test problems, we assess the model's capacity to generate code, documentation, and unit tests, as well as its ability to translate existing code between commonly used programming languages. Our comprehensive analysis evaluates the compilation, runtime behavior, and correctness of the generated and translated code. Additionally, we assess the quality of automatically generated code, documentation and unit tests. Our results indicate that while Llama 2-70B frequently generates syntactically correct and functional code for simpler numerical tasks, it encounters substantial difficulties with more complex, parallelized, or distributed computations, requiring considerable manual corrections. We identify key limitations and suggest areas for future improvements to better leverage AI-driven automation in scientific computing workflows.
\end{abstract}

\keywords{Large Language Models, LLama 2 70B, Software development, Programming languages}

\section{Introduction}

Large Language Models (LLMs) have made significant advances in various code-related tasks, particularly in generating source code from natural language descriptions (\cite{zhao2023survey, chang2024survey}). 
Their effectiveness is primarily driven by their extensive number of model parameters, the use of large and diverse datasets, and the immense computational resources employed during training (\cite{kaplan2020scaling}). These models are typically trained on vast corpora sourced from the web. LLMs are capable of capturing intricate patterns, linguistic subtleties, and semantic relationships. 

A wide range of models are available for code generation. There are general-purpose models like ChatGPT (\cite{ouyang2022training}), GPT-4 (\cite{achiam2023gpt}), and LLaMA (\cite{touvron2023llama}) which are designed for a broad range of applications, as well as specialized models such as StarCoder, Code LLaMA (\cite{roziere2023code}), DeepSeek-Coder, and Code Gemma that are optimized for code-related tasks. The integration of code generation with the latest advances in LLM technology is now an essential tool for many businesses, as well as an essential target for LLM developers as programming languages are considered to be different dialects of natural language (\cite{athiwaratkun2022multi}).



\subsection{Motivations}
Surveys conducted by PRACE~\cite{bull2011applications} on the usage of European supercomputers have shown that 
FORTRAN is the most popular programming language and \cpp\ is the second. Python is catching up, probably due to machine learning applications which are mostly written in Python~(\cite{hasheminezhad2020towards}). A survey from the National Energy Research Scientific Computing Center (NERSC) showed similar results.

Currently, most of the supercomputers in the Top\ 500 list are heterogeneous and make use of NVIDIA, AMD, or Intel GPUs. One can use CUDA FORTRAN for NVIDIA GPUs or ROCm FORTRAN for AMD GPUs, respectively. However, for Intel GPUs, the only tool is SYCL, and it does not provide a FORTRAN interface. The story is similar for Kokkos, a popular performance portability library for \cpp.

Therefore, to maximize the portability of scientific codes, efforts are being made to translate them from FORTRAN to \cpp~(\cite{grosse2012automatic}) or Python~(\cite{bysiek2016migrating}).


Another language that many wish to translate is \textbf{CO}mmon \textbf{B}usiness \textbf{O}riented \textbf{L}anguage (COBOL). This language is widely used in business, finance, and administrative systems. Because COBOL is less popular, COBOL programmers are harder to find. As a result, there is a desire to move away from the platform. Typically, the translation target for legacy COBOL code is Java~(\cite{sneed2010migrating}). IBM presented an AI-assisted COBOL to Java translation approach using IBM watsonx. 

Another commonly sought translation is the translation of MatLab code into R or Python. Here, the motivation is to move away from a commercial product and towards an open-source product.

For all of the above reasons and use cases, we believe it is useful to investigate the viability of translating code from one language to another using AI.

\subsection{Choice of Models}
LLaMA was first introduced by Meta as a collection of pretrained and fine-tuned LLM models, ranging from 7 billion to 70 billion parameters (\cite{touvron2023llama}, \cite{touvron2023llama2}). Code LLaMA (\cite{roziere2023code}), which was specifically designed for code generation, was initially released on August 24, 2023. This model is based on LLaMA 2 and has been fine-tuned for tasks focused on generating and comprehending code. For this paper, we have chosen \texttt{LLaMA 2-70B} (\cite{touvron2023llama2}) because of its large parameter count, which allows for enhanced performance in understanding and generating code. Its fine-tuning on code-specific datasets makes it particularly well-suited for handling a variety of programming tasks, including code generation, documentation, and translation.

Our analysis focuses on the compilation and correctness of the generated code. We selected programming languages from among the Top\ 15 languages in the TIOBE Index (\cite{tiobe}). The languages chosen for our analysis are listed in Table~\ref{tab:tiobe}. We picked \cpp and FORTRAN since these are used in High Performance computing. Python is used for artificial intelligence. Matlab and R used for data science. We included CUDA for GPU programming since this is one of the first languages and a large amount training data should be available.

The first example in this study is numerical integration; the second, a conjugate gradient solver that incorporates matrix and vector operations; and lastly, a parallel 1D stencil-based heat equation solver~(\cite{diehl2023benchmarking}). This last model problem was chosen due to its relevance in testing parallel computational workflows and the commonality of this type of code in scientific computing. 
%
\begin{table}[tb]
    \centering
    \begin{tabular}{c|ll}\toprule
     Rank & \multicolumn{2}{c}{Programming language}    \\\midrule
     1   &  \includegraphics[scale=0.1125]{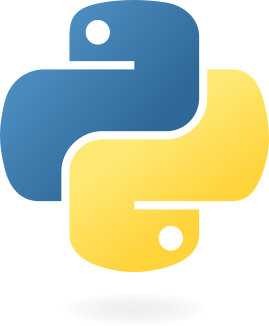} & Python \\
     2 &  \includegraphics[scale=0.025]{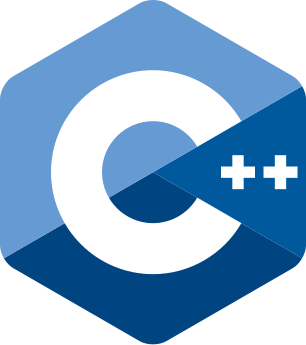} & \cpp \\
     4 &  \includegraphics[scale=0.025]{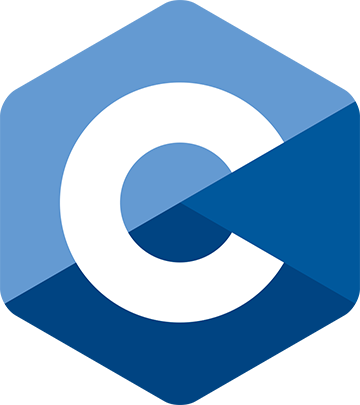} & C \\
     10 & \includegraphics[scale=0.025]{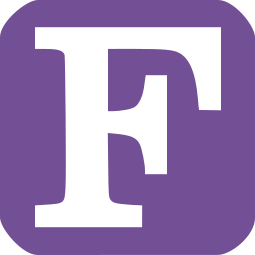} & FORTRAN \\
     12 & \includegraphics[scale=0.035]{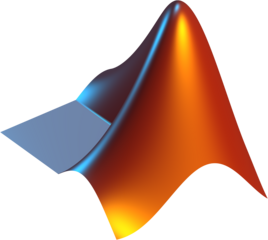} & Matlab \\
     15 & \includegraphics[scale=0.035]{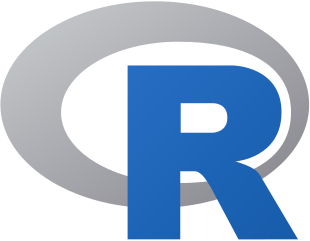} & R \\\bottomrule
    \end{tabular}
    \caption{Ranking of programming languages with respect to the \textit{Tiobe index} of September 2024. We picked \cpp and FORTRAN since these are used in High Performance computing. Python is used for artificial intelligence. Matlab and R used for data science. We included CUDA fir GPU programming since this is one of the first languages and most training data might be available (\cite{tiobe}).}
    \label{tab:tiobe}
\end{table}
%

\subsection{Paper Structure}
Section~\ref{sec:related:work} summarizes the related work. Section~\ref{sec:model:problems} briefly introduces the model problems used for the code generation. Section~\ref{sec:code:generation} presents the code generated by the AI, Section~\ref{sec:code:quality} analyzes the quality, and Section~\ref{sec:performance:generated} the scaling of the generated code. Section~\ref{sec:code:documentation} presents the generated documentation using the AI model. Section~\ref{sec:code:tests} presents the generated unit tests, and Section~\ref{sec:tests:quality} analyzes the quality of the unit tests. Section~\ref{sec:code:translated} presents the translated codes using the AI model, and Section~\ref{sec:translated:quality} analyzes the quality and Section~\ref{sec:code:performance:translated} the performance of the translated code. Finally, Section~\ref{sec:conclusion} concludes the work.

\section{Related work}
\label{sec:related:work} 
There is a large variety of programming benchmarks, with curated synthesis problems and test cases, used to measure
the performance of various LLMs on code generation.
HumanEval (\cite{chen2021evaluating}) is one of the pioneering and most widely studied human-written benchmarks for LLM-based code generation, consisting of 164 problems. Each provides a Python function signature and docstring and the associated test cases for correctness. MBPP (\cite{austin2021program}) is another Python-focused problem dataset comprising a set of 974 programming problems including the test cases. 
In multi-language coding scenarios, benchmarks such as HumanEval-X (\cite{zheng2023codegeex}) and MCoNaLa (\cite{wang2022mconala}) were used.

\noindent Burak et.\ al.~\cite{yeticstiren2023evaluating} compared the performance of three AI-code generation tools: GitHub Copilot, Amazon
CodeWhisperer, and OpenAI’s ChatGPT. The study evaluated the quality of the generated code in terms of validity, correctness,
 security, reliability, and maintainability. Their analysis reveals that the latest versions of ChatGPT, GitHub Copilot,
and Amazon CodeWhisperer generate correct code 65.2\%, 46.3\%, and 31.1\% of the
time, respectively.

\noindent More recently,~\cite{diehl2024evaluating} evaluated the capabilities of ChatGPT versions 3.5 and 4 in generating code across a
diverse range of programming languages taking into account compilation, runtime performance, and
accuracy of the codes. To this end, ChatGPT was asked to generate three distinct
codes: a simple numerical integration, a conjugate gradient solver, and a parallel 1D stencil-based
heat equation solver. Parallel codes---even the simple example we chose to study
here---are also difficult for the AI to generate correctly. 
\cite{liu2024your} propose EvalPlus, an evaluation framework to benchmark the functional correctness of LLM-generated code.  Their evaluation spanned 26 popular LLMs (\emph{e.g.}, GPT-4, ChatGPT, etc.). Additionally, \cite{cao2024javabench} proposed JavaBench, a project-level Java benchmark that exercises object-oriented programming features. It comprises four Java projects with 389 methods in 106 Java classes.

\noindent In the context of security, \cite{wang2024your} introduced CodeSecEval, a dataset targeting 44 critical vulnerability types with 180 samples, enabling automatic evaluation of code models in code generation and bug fixing tasks. Their findings reveal that current models often produce vulnerable code, overlooking security considerations. Similarly, in code translation, while LLMs achieve state-of-the-art performance in many tasks, they face challenges due to limited pretraining on parallel multilingual code. To address this, \cite{tao2024unraveling} proposed PolyHumanEval, a multilingual benchmark covering 14 programming languages to assess the ability of LLMs to translate between diverse programming language pairs.

\noindent Finally, \cite{ersoy2024benchmarking} benchmarks the capabilities of Llama 3 70B, for code generation tasks using HumanEval (\cite{chen2021evaluating}) and MBPP (\cite{austin2021program}). \cite{huang2024performance} indicate that Llama 2 holds significant promise for applications involving in-context learning, with notable strengths in both answer quality and inference speed.

\noindent In this paper, we focus on evaluating Llama 2 70B for code generation tasks, extending previous studies on LLM-based code synthesis. While prior work has demonstrated the potential of Llama 2 for in-context learning and its performance on benchmarks like HumanEval (\cite{chen2021evaluating}) and MBPP (\cite{austin2021program}), our evaluation provides a more focused and in-depth analysis of its capabilities in generating correct and efficient code. We specifically assess Llama 2 70B in various programming tasks, examining its performance in terms of code quality, scalability, translation, and the ability to handle complex and parallel code generation challenges.

\section{Model problems}
\label{sec:model:problems}

\subsection{Numerical integration}
\label{sec:model:problems:integration}
An example problem often used in undergraduate calculus textbooks is numerical integration using a Riemann sum to evaluate the area below the integral. We asked the AI model the following query:
\begin{displayquote}
\begin{tcolorbox}[colback=gray!15]

Write a \textbf{\{Python,\cpp,Fortran,Matlab,R\}} code to compute the area between -pi and 2/3pi for $\sin(x)$ and validate it.
\end{tcolorbox}
\end{displayquote}
to generate the Python, \cpp, FORTRAN, Matlab, and R code to integrate the area between $-\pi$ and $2/3\pi$ for the integral
\begin{align}
    \int\limits_{-\pi}^{2\pi/3} \sin(x) dx \text{.}
\end{align}
The correct result for the area using the fundamental theorem of calculus is given by
\begin{align}
    \int\limits_{-\pi}^{2\pi/3} \sin(x) = -\cos(2\pi/3) + \cos(-\pi) = -0.5 \text{.}
\end{align}
For Python, Matlab, and R, the AI generated code used a standard package to perform this task, however, for \cpp and FORTRAN it used a Riemann sum. 
\begin{displayquote}
\begin{tcolorbox}[colback=gray!15]

Write a \textbf{\{Python,Matlab,R\}} code to compute the area between -pi and 2/3pi for $\sin(x)$ using a Riemann sum and validate it.
\end{tcolorbox}
\end{displayquote}

For a fair comparison, we modified the query for the Python, Matlab, and R codes to request the Riemann sum instead. The codes were generated on 09/05/2024.

\subsection{Conjugate gradient}
\label{sec:model:problems:cg}
A more advanced example from numerical methods textbooks is to use a conjugate gradient solver to solve a system of linear equations
\begin{align}
    A^{n \times n} \cdot x^n = b^n \quad \text{with} \quad n \in \mathbb{Z}^+, A = A^t, \text{ and } x^t A z > 0, \forall X \in \mathbb{R}^n\text{.} 
\end{align}
For more detail, we refer the reader to this excellent introduction~(\cite{shewchuk1994introduction}).  For this task, we gave the AI model the following query:
\begin{displayquote}
\begin{tcolorbox}[colback=gray!15]

Write a \textbf{\{Python, \cpp, Fortran, Matlab, R\}} code to solve the linear equation system using the conjugate gradient solver and validate it. 
\end{tcolorbox}
\end{displayquote}
To evaluate the code, we wrote a program to use the generated solver on the following system of equations:
\begin{align}
M \cdot x = b \quad \text{with}\quad A = \left( \begin{matrix} 4 & -1 & 0 \\ -1 & 4 & -1 \\ 0 & -1 & 4 \end{matrix} \right) \text{ and }  b = \left( \begin{matrix} 1 \\ 2 \\ 3  \end{matrix} \right) \text{.}
\end{align}
The solution $x$ is $(0.46428571, \;0.85714286, \;0.96428571)^T$. The codes were generated on 09/09/2024.

\subsection{Heat equation solver}
\label{sec:model:problems:heat}
In the previous examples, we focused on the generation of single-threaded code. Now, the focus is on the generation of parallel and distributed code. For the distributed code, we only asked the model to generate a Python, FORTRAN, and \cpp\ code. Specifically, we requested a one-dimensional heat equation solver in this study. The one-dimensional heat equation query reads as follows:
\begin{align}
    \frac{\partial u}{\partial  t} = \alpha\frac{\partial^2 u}{\partial x^2}, \quad 0 \leq c < L, t >0
\end{align}
where $\alpha$ is the material's diffusivity. For the discretization in space a finite different scheme
\begin{align}
    u(x_i,t+1) = u(x_i,t) + dt \alpha \frac{u(x_{i-1},t) - 2 u(t,x_i) - u(t,x_{i+1})}{h^2}
\end{align}
We requested that the Euler time integration should be used. The grid was not specified, but we expected the AI to choose an equidistant linear grid with $n$ grid points $x = \{ x_i = i \cdot h \in \mathbb{R} \vert i = 0,1\ldots,n-1 \}$. We asked the AI model the following query

\begin{displayquote}
\begin{tcolorbox}[colback=gray!15]

Write a \textbf{\{parallel, distributed\}} \textbf{\{Python,\cpp,Fortran,Matlab,R\}} code to solve the one-dimensional heat equation using a finite difference scheme for the discretization in space and the Euler method for time integration and validate it.
\end{tcolorbox}

\end{displayquote}
To test the Python, \cpp, FORTRAN, Matlab, and R code. We initialized $u(0,x) = \sin(\pi  x)$ and set $\alpha=0.1$, $n=100$, $h= L/ (n-1)$, $L=1$, $dt=T/1000$, and $T=0.1$. The analytic solution for the final distribution at time $T$ using the above parameters is given as
\begin{align}
    u(T,x) = \sin(\pi  x) \cdot \exp(-\alpha \pi^2  t) \text{.}
\end{align}
The parallel codes were generated on 09/11/2024. The AI model used the Message Passing Interface (MPI) for both queries. One might assume that the AI model should use OpenMP for the \textbf{parallel} code and MPI for the \textbf{distributed} code.

\begin{displayquote}
\begin{tcolorbox}[colback=gray!15]

Write a \textbf{\{parallel\}} \textbf{\{\cpp,Fortran\}} code using OpenMP to solve the one-dimensional heat equation using a finite difference scheme for discretization in space and the Euler method for time integration and validate it.
\end{tcolorbox}

\end{displayquote}

Therefore, we asked on 09/16/2024 to generate a parallel code using OpenMP.

The queries for the CUDA codes were generic (\emph{i.e.}\ we did not specify a specific GPU architecture) all started with "Write a CUDA code". Indeed, we have tried to include a specific architecture (\emph{e.g.}\ Ampere, Hopper) in the query, but this did not change the output of the AI model.

In the following sections, we discuss the codes generated by the model. For each section, we provide a single example. All the other codes are available on our public repository\footnote{\url{https://github.com/diehlpkpapers/ai-journal-paper}}. 

\section{Code generation}
\label{sec:code:generation}

\subsection{Python}


\textbf{Numerical Integration. }
Listing~\ref{python:integration:scipy} shows the Python code generated for the numerical integration. The AI used \lstinline[language=python]{numpy} for the \lstinline{np.sin} function and \lstinline[language=python]{np.pi} for the constant $\pi$. The package \lstinline[language=python]{scipy} was used for the integration. 
\begin{displayquote}
\begin{tcolorbox}[colback=gray!15]

Write a \textbf{\{Python\}} code to compute the area between -pi and 2/3pi for $\sin(x)$ and validate it without using scipy.
\end{tcolorbox}
\end{displayquote}

\begin{figure}[tb]
    \centering
    \includegraphics[width=\linewidth]{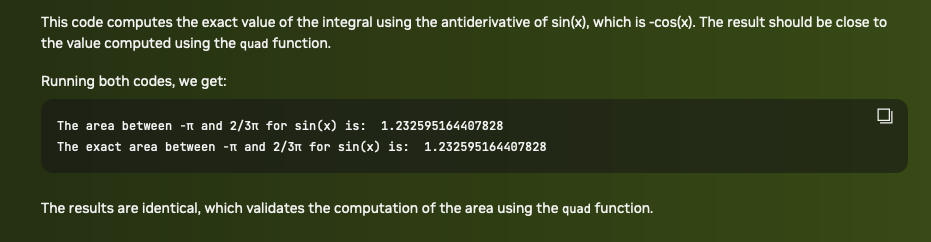}
    \caption{Output for the evaluation of the generated Python models by the AI model. Obviously both numbers are incorrect, however, by executing the generated code, the correct result is printed.}
    \label{fig:python:eval}
\end{figure}

To make a fair comparison with Fortran and \cpp, we asked the AI to generate a Python version without using \lstinline[language=python]{scipy}, depicted in Listing~\ref{python:integration:riemann}. After that, the AI generated plain Python code used a Riemann sum. The AI model mentioned that the parameter $n$ of the \lstinline{for}loop controls the accuracy of the result. For the validation of the generated codes, the AI model recommended that we compare its answer with the exact value of the integral, which can be computed using the fundamental theorem of calculus $-np.cos(-\pi) + np.cos(2/3\pi)$. The AI model then generated a Python code to compute the exact solution. This code was similar to the first generated code and the line \lstinline[language=python]{from scipy.integrate import quad} was removed and the computation of the result was replaced by \lstinline{exact_result = -np.cos(b) + np.cos(a)}. Figure~\ref{fig:python:eval} shows the solution of both codes. It then claims that since both values are the same, the generated code is correct. It is obvious that the results are wrong, however, running both codes give the correct results. \change{For ChatGPT 4, in the author's previous work~\cite{diehl2024evaluating}, the generated Python code had runtime errors and the result was not correct. For the DeepSeek model, the code was executed and produced correct results.}

%
%

\begin{lstlisting}[language=python,caption=Generated python code for the numerical integration using scipy.,label=python:integration:scipy,escapechar=|, float=false]
import numpy as np
from scipy.integrate import quad

# define the function sin(x)
def f(x):
    return np.sin(x)

# define the limits of integration
a = -np.pi
b = 2*np.pi/3

# compute the integral using quad
result, error = quad(f, a, b)

print("The area between -π and 2/3π for sin(x) is: ", result)
\end{lstlisting}

\begin{lstlisting}[language=python,caption=Generated python code for the numerical integration using a Riemann sum.,label=python:integration:riemann,escapechar=|, float=!htbp]
import numpy as np

# define the function sin(x)
def f(x):
    return np.sin(x)

# define the limits of integration
a = -np.pi
b = 2*np.pi/3

# define the number of intervals for the Riemann sum
n = 1000

# compute the width of each interval
dx = (b - a) / n

# initialize the sum
sum = 0

# compute the Riemann sum
for i in range(n):
    x = a + i * dx
    sum += f(x) * dx

print("The area between -π and 2/3π for sin(x) is: ", sum)
\end{lstlisting}

\newpara
\noindent\textbf{Conjugate Gradient solver} 
The generated python code uses the \lstinline[language=python]{numpy} package for the matrix and vector data structure. A function \lstinline[language=python]{conjugate_gradient} is defined containing the algorithm for the conjugate gradient solver. The result of the solver in the function is compared with the built-in solver \lstinline[language=python]{np.linalg.solve}. The function \lstinline[language=python]{np.testing.assert_allclose} is used to compare the two solution vectors. The code executed and showed the correct results. The AI model mentioned that the variable \lstinline[language=python]{tol} and \lstinline[language=python]{max_iter} control the convergence of the solver.

\begin{figure}[tb]
    \centering
    \includegraphics[width=0.75\linewidth]{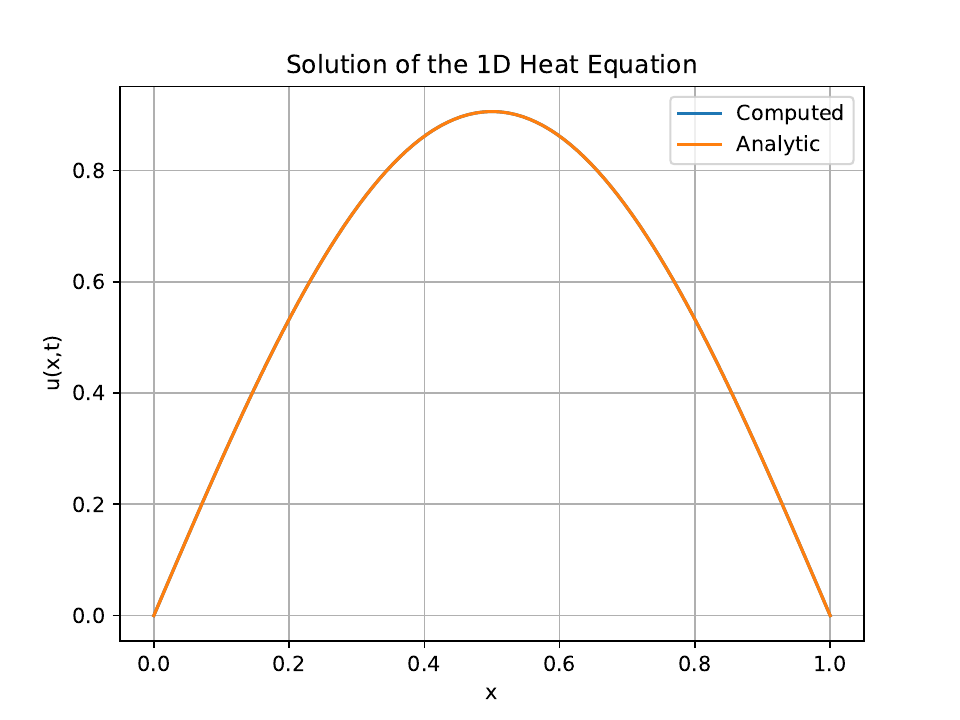}
    \caption{Plot of the analytic solution and the computed solution by the Python code. The AI model added the functionality for the plot to the code using Python's Matplotlib. We changed the layout by adding grid lines and saved it as a PDF document instead of just showing the plot. }
    \label{fig:python:heat:plot}
\end{figure}

\newpara
\noindent\textbf{Parallel Heat equation solver}
The AI model used Python's multiprocessing tool (\lstinline[language=python]{import multiprocessing as mp}). It correctly generated a loop to iterate over the nodes and apply the stencil for the finite difference scheme, however, the code did not execute because the variable for the time step size \lstinline[language=python]{dt} was not declared. After a declaration of the variable was added, the code executed and produced the correct results. The AI model added a function to compute the analytic solution and some functionality to plot the analytic function and the computed results using Python's Matplotlib. Figure~\ref{fig:python:heat:plot} shows the generated plot.

\subsection{\cpp}

\textbf{Numerical Integration. } 
The generated \cpp code uses the \lstinline[language=c++]{#include <cmath>} header for the $\sin()$ function and the \lstinline[language=c++]{M_PI} constant for $\pi$. The code used the Riemann sum to approximate the area. Again, the code used the fundamental theorem of calculus $-np.cos(-\pi) + np.cos(2/3\pi)$ to check the result of the Riemann sum. The AI model provided the same information as for the Python-generated code. For example, the AI model mentioned that the variable $n$ controls the accuracy of the result. The generated code compiled, executed, and produced the correct result. However, the AI model showed the wrong results for both codes again. \change{For both ChatGPT 4 and DeepSeek, in the author's previous work~\cite{diehl2024evaluating}, the generated code compiled, executed, and produced correct results.}

\newpara
\noindent
\textbf{Conjugate Gradient solver} 
The code uses \lstinline[language=c++]{std::vector<double>} for the vector and \lstinline[language=c++]{std::vector<std::vector<double>>} for the matrix. The code compiled using a recent \cpp\ standard, executed successfully and produced the correct result. The code used Gaussian elimination to validate the result.

\newpara
\noindent
\textbf{Parallel Heat equation solver} The generated \cpp code for the heat equation in parallel compiled, executed, and produced the correct result. The code included the header file \lstinline[language=c++]{#include <omp.h>} which was not necessary since only an OpenMP pragma was used and no functions of this header were called. We had to change the final time from 1 to 0.1 to align with the parameters in the other test problem. The AI model used \lstinline[language=c++]{#pragma omp parallel for private(i)} to execute the loop to update the temperature. However, the \lstinline[language=c++]{private(i)} is not necessary and removing it does not affect the correctness of the code.

\subsection{FORTRAN}
 
\textbf{Numerical Integration. }
The generated code did not use a constant value from the standard library or math library for $\pi$ and used $3.14159265359$ as an approximation of $\pi$. This is due to the fact that FORTRAN has no built-in constant for \lstinline[language = c]{PI}. The code used the Riemann sum to compute the area and the AI model mentioned that the parameter $n$ controls the accuracy of the result. The code compiled, executed, and produced the correct result. \change{For ChatGPT 4, in the author's previous work~\cite{diehl2024evaluating}, the generated code compiled and executed, however, it did not produce correct results. For the DeepSeek model, the code compiled, executed, and produced correct results.}

\newpara
\noindent
\textbf{Conjugate Gradient solver} 
The generated FORTRAN code did not compile. There were undeclared variables. For some data types \lstinline[language=fortran]{INTENT(INT)} was declared, however, the types were updated within the scope and the data types should be declared as \lstinline[language=fortran]{INTENT(OUT)}. Once these errors were corrected and the code was compiled, the code executed, but did not produce the correct results. The issue was a problem with the computation of the residual. After fixing that, the code produced the correct results. 


\newpara
\noindent
\textbf{Parallel Heat equation solver}  The generated code did not compile because the variable for $\pi$ was not declared. In addition, the loop variables were not declared. The AI model used \lstinline[language=fortran]{!\$OMP PARALLEL DO PRIVATE(i)} to execute the loop to update the temperature. However, the \lstinline[language=c++]{PRIVATE(i)} was not necessary and removing it did not affect the correctness of the code. After fixing these issues, the code compiled. However, all final values of the temperature were \lstinline[language=fortran]{NaN}. This was due to the choice of the parameters and the simulation blow up. Using the parameters of the model problem produced the correct results.

For the distributed code the variable for $\pi$ was not declared. After declaration the code compiled and executed.

\subsection{Matlab}

\textbf{Numerical Integration. }
The generated MATLAB code for the numerical example used the \lstinline[language=matlab]{trapz} function. However, upon execution, an error occurred: \texttt{Unrecognized function or variable 'x'} indicating a problem with variable \lstinline[language=matlab]{x} initialization. The authors manually fixed the code by replacing \lstinline[language=matlab]{trapz} with the \lstinline[language=matlab]{integral} function, which successfully provided the correct output. \change{ The code generated by GPT-4 executed successfully bit did not produce correct results (\cite{diehl2024evaluating}). In contrast, DeepSeek produced valid MATLAB code that executed correctly, and produced accurate results.}

\newpara \noindent
\textbf{Conjugate Gradient solver} 
 The generated MATLAB code defined a linear system of equations and solved it using the conjugate gradient method. The solution was compared with the MATLAB built-in solver \lstinline[language=matlab]{A\b}. The code executed correctly, and the comparison between the custom CG solver and the built-in solver showed consistent results. \change{Comparing to our previous results (\cite{diehl2024evaluating}), the MATLAB code generated by GPT-4 executed successfully, but no valid output was obtained. In contrast, DeepSeek generated correct MATLAB code for this task.}

\newpara \noindent
\textbf{Parallel Heat equation solver} The LLM model used the parpool package (via \lstinline[language=matlab]{parpool(num_workers)}). However, the code did not execute because the variable for the time step size, \lstinline[language=matlab]{dt}, was not declared. After declaring the variable, the code executed but did not produce correct results. The AI model also added a function to compute the exact solution and functionality to plot both the analytic and computed results. At first, the executed code plotted all the computed solutions at each step. The authors changed the code and plotted only the final computed solution. The generated plots for the computed and exact solutions clearly illustrate that the computed output is incorrect is on the github repository \footnote{\url{https://github.com/diehlpkpapers/ai-journal-paper/blob/main/matlab/HeatComputedSolutionMatlab.png}}. The numerical solution deviates from the exact solution, this indicate potential issues with boundary conditions. \change{In contrast, the code generated by GPT-4 executed successfully and produced correct results (See \cite{diehl2024evaluating}). The code generated by DeepSeek, however, had runtime errors; however; after fixing the error correct results are produced.}




\subsection{R}

\textbf{Numerical Integration.}
When testing the AI-generated codes in R, the code generated \change{by LLamA} used the \lstinline[language =R]{Riemann sum} method, executed correctly, and produced accurate results. \change{In contrast, the code generated by GPT-4 did not produce correct results (\cite{diehl2024evaluating}). The code generated by DeepSeek initially contained errors related to the \lstinline[language=R]{sprintf} function; however; after fixing the error, correct results were obtained.}

\newpara
\noindent
\textbf{Conjugate Gradient solver. } 
 The generated code used the function \lstinline[language=R]{conjugate_gradient}, which iteratively updates the solution vector \lstinline[language=R]{x}, the residual \lstinline[language=R]{r}, and the search direction \lstinline[language=R]{p}. After obtaining the solution using the CG solver, the result was compared with the solution generated by the built-in \lstinline[language=R]{solve} function. The code executed successfully, with the CG solver producing results consistent with the exact solution. \change{DeepSeek produced similarly accurate results after execution. In contrast, the code generated by GPT-4 ran without errors, but the output was not consistent with the exact solution (\cite{diehl2024evaluating}).}

\newpara
\noindent
\textbf{Parallel Heat equation solver} 
Here, the script generated by LlamA used the \lstinline[language =R]{library(parallel)}. However, the code did not execute because the variable \lstinline[language =R]{dt} was not defined. After fixing the error, the code executed but did not produced correct result.
This numerical diffusion may be due to incorrect boundary conditions, which prevents the solution from converging to the exact result. The generated plots for the computed and exact solutions, clearly illustrate that the computed output is incorrect is on the github repository \footnote{\url{https://github.com/diehlpkpapers/ai-journal-paper/blob/main/r/HeatComputedSolution.png}}. \change{When comparing to other models in our previous work, the code generated by GPT-4 executed without errors and produced correct results (\cite{diehl2024evaluating}). In the case of DeepSeek, the initial version failed to include a required package \lstinline[language =R]{library(parallel)}. After adding the required package include, the code executed successfully and produced correct results.}

\subsection{CUDA}

\textbf{Numerical Integration. } Similar to the previous examples, the AI generated CUDA implementation used  the \lstinline[language=c++]{M_PI} constant for $\pi$ and \lstinline[language=c++]{#include <cmath>} for calling the $\sin()$ function inside the kernel. It also used \lstinline[language=c++]{atomicAdd} to update the shared variable in the global memory, which correctly handles multiple concurrent accesses. The generated code compiled and executed but did not produce the correct result. Instead of writing a parallel code, it generates a code where each thread redundantly computes the entire problem. In addition, the generated analytical solution is also incorrect because the code used a wrong formula (\emph{i.e.}\ $ 2 - \frac{\sqrt(3)}{2} $) to compute the exact solution for validation. 

\newpara
\noindent
\textbf{Conjugate Gradient solver} The generated code compiled and executed but did not produce the correct result. It also contained multiple memory management bugs that triggered runtime errors. It seems that the AI model has difficulty in differentiating the host from the device memory. For example, it tries to copy data from the host to the device by passing two device pointers, or it tries to access host memory from a CUDA kernel. The code is hard to read due to multiple inconsistencies in variable naming. For example, it respects a naming convention in the \textit{main}, preceding the device memory pointers by \textit{"d\_"}, but omits it when those virtual pointers are passed to the \lstinline[language=c++]{conjugateGradient()} function. 

\newpara
\noindent
\textbf{Parallel Heat equation solver} 
The generated CUDA code compiled but was not able to produce a correct result because the input parameters were not chosen correctly, resulting in numerical instability. Also, the code did not properly check the boundary conditions, which resulted in illegal memory accesses. After fixing these issues, the generated CUDA code compiled, executed, and produced the correct result. Also, the model produced a correct validation test, using the appropriate analytical solution.

\subsection{MPI}

\textbf{\cpp} The generated distributed \cpp code compiled and executed but did not produce the correct results. The code correctly called \textit{MPI\_init()} and \textit{MPI\_finalize()}, but did not trigger any communication. The only invocation of the communication library was for creating a contiguous type that was never used. As a result, the generated code was serial and did not make use of a distributed system. The serial code itself also contained several mistakes, such as allocating and freeing memory every time step or not taking into account the previous iterations when computing a new value.

\newpara
\noindent
\textbf{Python} The generated code used mpi4py for leveraging the MPI communication library. It used individual \textit{send} and \textit{recv} operations for the halo exchange, which in a real code would be probably replaced by `sendrecv` operations. It run successfully with two processes but did not produce the correct result. With more than 3 processes, the code produced a deadlock due to an incorrect communication workflow (\emph{e.g.}\ some processes wait on receiving messages that are never sent). Besides the deadlock, the code contained numerous errors: incorrect initialization, not handling cases where the number of process is not divisible by the problem size, not reserving additional cells in the array for storing the ghost cells, and sending the wrong data. We have decided not to offer a fix for the code, as this would mean rewriting it almost entirely.

\newpara
\noindent 
\textbf{Fortran} The generated code compiled, executed, and (after fixing the input parameters) even produced the correct result. However, similarly to the \cpp distributed code, the Fortran code was not different from a serial implementation. Indeed, the code invoked a routine from the MPI library exactly once. At the beginning then process zero initialized the entire domain and broadcasted it to all other processes. The rest of the code is equivalent to a serial program where each individual process computed the entire domain. 

\begin{table}[tbp]
    \centering
    \rowcolors{2}{gray!25}{white}
    \begin{tabular}{l|cccccc}\toprule
    Model & \multicolumn{2}{c}{Llama2} & \multicolumn{2}{c}{\textcolor{azure}{ChatGPT 4}} &  \multicolumn{2}{c}{\textcolor{darkgreen}{DeepSeek}}  \\\midrule
    Language   & \cpp & FORTRAN & Python & Matlab & R & CUDA \\\midrule
    Attribute & \multicolumn{6}{c}{Numerical integration} \\\midrule  
     Compilation    &  \checkmark / \textcolor{azure}{\checkmark} / \textcolor{darkgreen}{\checkmark}  & \checkmark / \textcolor{azure}{\checkmark} /\textcolor{darkgreen}{\checkmark}   & -- / \textcolor{azure}{--} / \textcolor{darkgreen}{--} & -- / \textcolor{azure}{--} / \textcolor{darkgreen}{--} & -- / \textcolor{azure}{--} / \textcolor{darkgreen}{--} &  \checkmark \\
     Runtime    &  \checkmark / \textcolor{azure}{\checkmark} / \textcolor{darkgreen}{\checkmark} & \checkmark / \textcolor{azure}{\checkmark} /\textcolor{darkgreen}{\checkmark} & \checkmark / \textcolor{azure}{\tikzxmark} / \textcolor{darkgreen}{\checkmark} &  \tikzxmark \, /  \textcolor{azure}{\checkmark} /  \textcolor{darkgreen}{\checkmark} & \checkmark / \textcolor{azure}{\checkmark}/ \textcolor{darkgreen}{\tikzxmark}  & \checkmark \\
     Correctness   &  \checkmark / \textcolor{azure}{\checkmark} / \textcolor{darkgreen}{\checkmark} & \checkmark / \textcolor{azure}{\tikzxmark} / \textcolor{darkgreen}{\checkmark}& \checkmark / \textcolor{azure}{\tikzxmark} / \textcolor{darkgreen}{\checkmark} & \tikzxmark \, / \textcolor{azure}{\tikzxmark}/  \textcolor{darkgreen}{\checkmark} & \checkmark / \textcolor{azure}{\tikzxmark}/ \textcolor{darkgreen}{\checkmark} & \tikzxmark \\\midrule
   & \multicolumn{6}{c}{Conjugate gradient} \\\midrule  
    Compilation    & \checkmark / \textcolor{azure}{\checkmark} / \textcolor{darkgreen}{\checkmark}   & \tikzxmark / \textcolor{azure}{\checkmark} / \textcolor{darkgreen}{\tikzxmark}  & -- / \textcolor{azure}{--} / \textcolor{darkgreen}{--} & -- / \textcolor{azure}{--} / \textcolor{darkgreen}{--} & -- / \textcolor{azure}{--} / \textcolor{darkgreen}{--} & \checkmark\\
     Runtime    & \checkmark / \textcolor{azure}{\checkmark} / \textcolor{darkgreen}{\checkmark} & \checkmark / \textcolor{azure}{\checkmark} / \textcolor{darkgreen}{\checkmark}  & \checkmark / \textcolor{azure}{\checkmark} / \textcolor{darkgreen}{\checkmark} & \checkmark / \textcolor{azure}{\checkmark}/\textcolor{darkgreen}{\checkmark}  &\checkmark / \textcolor{azure}{\checkmark}/\textcolor{darkgreen}{\checkmark} & \tikzxmark\\
     Correctness   & \checkmark / \textcolor{azure}{\checkmark} / \textcolor{darkgreen}{\checkmark} & \tikzxmark / \textcolor{azure}{\checkmark} / \textcolor{darkgreen}{\checkmark} & \checkmark / \textcolor{azure}{\checkmark} / \textcolor{darkgreen}{\checkmark} & \checkmark / \textcolor{azure}{\tikzxmark}/\textcolor{darkgreen}{\checkmark}  & \checkmark / \textcolor{azure}{\tikzxmark}/\textcolor{darkgreen}{\checkmark} & \tikzxmark \\\midrule
        & \multicolumn{6}{c}{Parallel heat equation solver} \\\midrule  
    Compilation    & \checkmark / \textcolor{azure}{\tikzxmark} / \textcolor{darkgreen}{\checkmark}   & \tikzxmark / \textcolor{azure}{\checkmark} / \textcolor{darkgreen}{\tikzxmark}   & -- / \textcolor{azure}{--} / \textcolor{darkgreen}{--} & -- / \textcolor{azure}{--} / \textcolor{darkgreen}{--} & -- / \textcolor{azure}{--} / \textcolor{darkgreen}{--} & \checkmark \\
     Runtime    & \checkmark / \textcolor{azure}{\checkmark} / \textcolor{darkgreen}{\checkmark} &  \checkmark / \textcolor{azure}{\checkmark} / \textcolor{darkgreen}{\checkmark} & \tikzxmark / \textcolor{azure}{\tikzxmark} / \textcolor{darkgreen}{\checkmark} &  \tikzxmark / \textcolor{azure}{\checkmark} / \textcolor{darkgreen}{\tikzxmark} & \tikzxmark / \textcolor{azure}{\checkmark}/ \textcolor{darkgreen}{\tikzxmark} & \tikzxmark \\
     Correctness   & \checkmark / \textcolor{azure}{\checkmark} / \textcolor{darkgreen}{\checkmark} & \checkmark / \textcolor{azure}{\tikzxmark} / \textcolor{darkgreen}{\checkmark}  & \checkmark / \textcolor{azure}{\checkmark} / \textcolor{darkgreen}{\checkmark} &  \tikzxmark / \textcolor{azure}{\checkmark} /\textcolor{darkgreen}{\checkmark} & \tikzxmark / \textcolor{azure}{\checkmark}/ \textcolor{darkgreen}{\checkmark} & \checkmark \\\bottomrule
    \end{tabular}
    \caption{Results for serial codes for numerical integration and conjugate gradient solver. For the parallel heat equation solver the AI model generated MPI code and we asked in a second query to generate a code using OpenMP. For the \cpp, FORTRAN, and CUDA codes, we check if the codes compile with the compilers shown in Table~\ref{tab:code:version}. For all codes, we check if the codes executed without any run time error, like index out of bound exceptions, and finally we checked if the code produces the correct results for the test cases in Section~\ref{sec:model:problems}. The results for ChatGPT 4 and DeepSeek were taken from the authors other works~\cite{diehl2024evaluating} and \cite{nader2025llmhpcbenchmarkingdeepseeks}, respectively.}
    \label{tab:code:results}
\end{table}

\section{Code quality}
\label{sec:code:quality}
Table~\ref{tab:code:results} shows the results for all three test problems. We use the following criteria to check the quality of the codes
\begin{itemize}
    \item Compilation:\\
    We saved the generated code to a text file and compiled the codes with recent compilers (See Table~\ref{tab:code:version}). If the compiler stopped with any error, we updated the code until it compiled. Note that Python, Matlab, and R codes are interpreted languages and so this step did not apply.
    \item Runtime:\\
    We executed the compiled codes or ran the interpreted codes. We checked whether the code finished without any errors, \emph{e.g.}\ an index out of bounds exception. For this criterion, We did not check the correctness of the results, only whether it finished. If we experienced issues with the code, we debugged the code and fixed the bugs.
    \item Correctness: 
    We checked whether the code produced the correct results using the input parameters and solutions in Section~\ref{sec:model:problems}. For some codes, we had to update the parameters proposed by the AI model to the ones matching our examples.    
\end{itemize}
\change{For the comparison of different LLMs, we added the results obtained by ChatGPT\ 4 from the author's previous work~\cite{diehl2024evaluating}. For the DeepSeek model, we added the conjugate gradient and parallel heat equation solver for \cpp, FORTRAN, and Python from another of the author's previous works~\cite{nader2025llmhpcbenchmarkingdeepseeks}. The focus of the paper was on HPC, therefore, the numerical integration example, as well as the Matlab and R examples, were not studied. We generated the code for the missing examples and languages on 05/14/2025.}

In the first third of the table, the results for the numerical integration in Section~\ref{sec:model:problems:integration} are shown. The \cpp and FORTRAN codes were both compiled using the GNU 12 compiler. None of the codes, except for Matlab, had runtime errors and produced the correct result. Thus, for this relatively easy example, the AI model could generate codes. However, it is not clear why the Matlab generated code had issues.

In the second third of the table, the results for the conjugate gradient solver in Section~\ref{sec:model:problems:cg} are shown. The \cpp code compiled, but the FORTRAN code did not due to incorrect \texttt{parameter} declarations on some variables. All generated codes executed. Except for the FORTRAN code, all codes produced the correct results. For the FORTRAN code, the computation of the residual was wrong, however, after fixing that to the correct equation, the code produced the correct results. The last third of the table shows the parallel implementation of the heat equation solver in Section~\ref{sec:model:problems:heat}. Note that the AI model generated code using the Message Passing Interface (MPI) and we extended the query to ask for a parallel code using OpenMP.

Table~\ref{tab:code:results_distributed} shows the results for the distributed code in \cpp, FORTRAN and Python. We excluded  Matlab and R code, since it is not common to write distributed codes using these programming languages. Table~\ref{tab:code:version} lists the version of all used tools to evaluate the generated code.

\begin{table}[tbp]
    \centering
    \rowcolors{2}{gray!25}{white}
    \begin{tabular}{l|ccc}\toprule
    Language     & \cpp & FORTRAN  & Python   \\\midrule
     Compilation & \checkmark  & \checkmark & -  \\
     Runtime     & \tikzxmark & \tikzxmark & \tikzxmark  \\
     Correctness & \tikzxmark & \checkmark & \tikzxmark  \\\bottomrule
    \end{tabular}
    \caption{Results for distributed codes using the Message Passing Interface (MPI). We check if the codes compile with the compilers shown in Table~\ref{tab:code:version}. For all codes, we check if the codes executed without any run time error, like index out of bound exceptions, and finally we checked if the code produces the correct results for the test cases in Section~\ref{sec:model:problems}.}
    \label{tab:code:results_distributed}
\end{table}

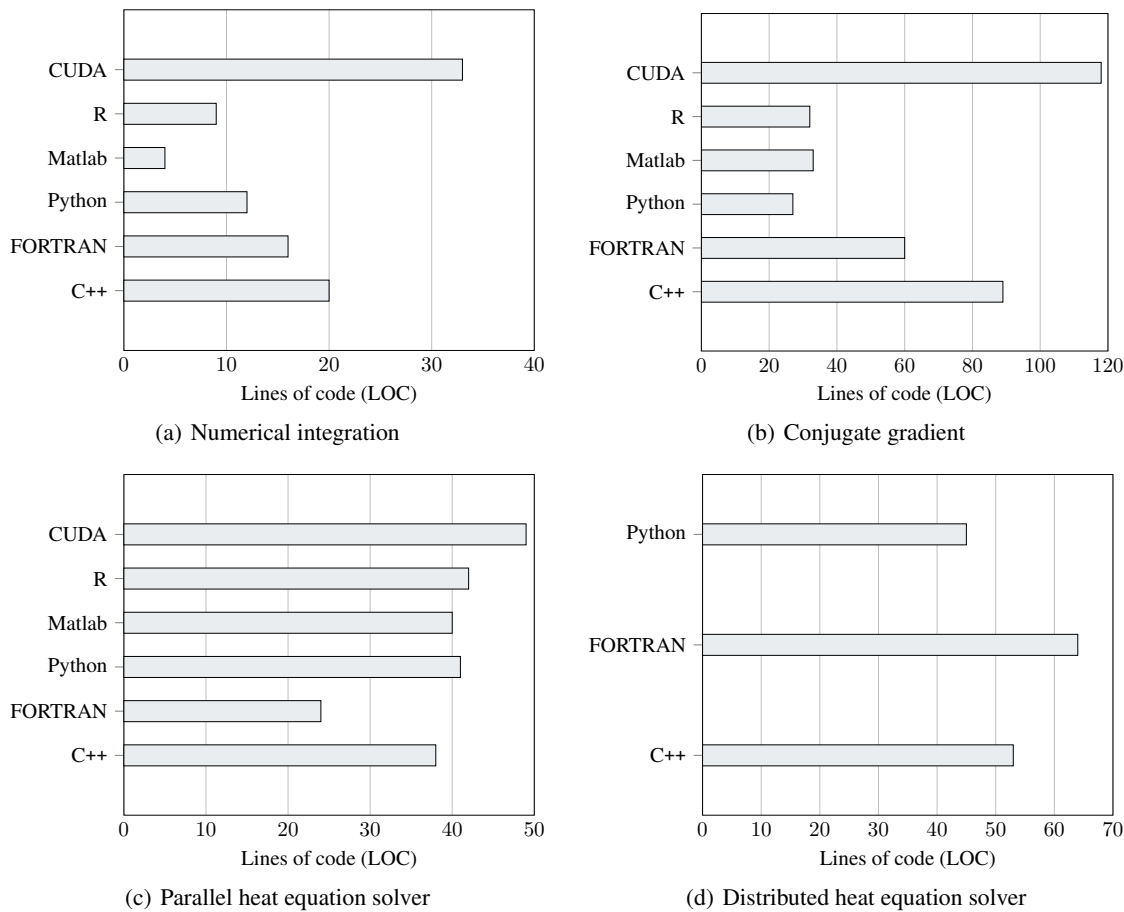
\begin{figure}[tb]
    \centering

\subfigure[Numerical integration\label{fig:code:lines:of:code:numerical}]{
\resizebox{0.45\textwidth}{!}
        {
\begin{tikzpicture}
 \begin{axis}[
    xbar=12pt,
    xmin=0,xmax=40,
    ytick=data,
    enlarge y limits={abs=1cm},
    symbolic y coords={C\texttt{++},FORTRAN,Python,Matlab,R,CUDA},
    bar width = 10pt,
    xlabel= Lines of code (LOC), 
    ytick align=outside, 
    ytick pos=left,
    major x tick style ={ transparent},
    legend style={at={(0.04,0.96)},anchor=north west, font=\footnotesize, legend cell align=left},
    xmajorgrids=true
        ]    
    \addplot[xbar,fill=cadetgrey!20, area legend] coordinates {
        (20,C\texttt{++})
        (16,FORTRAN)
        (12,Python)
        (4,Matlab)
        (9,R)
        (33,CUDA)
        };
\end{axis}
\end{tikzpicture}
}
}
\subfigure[Conjugate gradient\label{fig:code:lines:of:code:cg}]{
\resizebox{0.45\textwidth}{!}
        {
\begin{tikzpicture}
 \begin{axis}[
    xbar=12pt,
    xmin=0,xmax=120,
    ytick=data,
    enlarge y limits={abs=1cm},
    symbolic y coords={C\texttt{++},FORTRAN,Python,Matlab,R,CUDA},
    bar width = 10pt,
    xlabel= Lines of code (LOC), 
    ytick align=outside, 
    ytick pos=left,
    major x tick style ={ transparent},
    legend style={at={(0.04,0.96)},anchor=north west, font=\footnotesize, legend cell align=left},
    xmajorgrids=true
        ]    
    \addplot[xbar,fill=cadetgrey!20, area legend] coordinates {
        (89,C\texttt{++})
        (60,FORTRAN)
        (27,Python)
        (33,Matlab)
        (32,R)
        (118,CUDA)
        };
\end{axis}
\end{tikzpicture}
}
}

\subfigure[Parallel heat equation solver\label{fig:code:lines:of:code:heat:parallel}]{
\resizebox{0.45\textwidth}{!}
        {
\begin{tikzpicture}
 \begin{axis}[
    xbar=12pt,
    xmin=0,xmax=50,
    ytick=data,
    enlarge y limits={abs=1cm},
    symbolic y coords={C\texttt{++},FORTRAN,Python,Matlab,R,CUDA},
    bar width = 10pt,
    xlabel= Lines of code (LOC), 
    ytick align=outside, 
    ytick pos=left,
    major x tick style ={ transparent},
    legend style={at={(0.04,0.96)},anchor=north west, font=\footnotesize, legend cell align=left},
    xmajorgrids=true
        ]    
    \addplot[xbar,fill=cadetgrey!20, area legend] coordinates {
        (38,C\texttt{++})
        (24,FORTRAN)
        (41,Python)
        (40,Matlab)
        (42,R)
        (49,CUDA)
        };
\end{axis}
\end{tikzpicture}
}
}
\subfigure[Distributed heat equation solver\label{fig:code:lines:of:code:heat:distributed}]{
\resizebox{0.45\textwidth}{!}
        {
\begin{tikzpicture}
 \begin{axis}[
    xbar=12pt,
    xmin=0,xmax=70,
    ytick=data,
    enlarge y limits={abs=1cm},
    symbolic y coords={C\texttt{++},FORTRAN,Python},
    bar width = 10pt,
    xlabel= Lines of code (LOC), 
    ytick align=outside, 
    ytick pos=left,
    major x tick style ={ transparent},
    legend style={at={(0.04,0.96)},anchor=north west, font=\footnotesize, legend cell align=left},
    xmajorgrids=true
        ]    
    \addplot[xbar,fill=cadetgrey!20, area legend] coordinates {
        (53,C\texttt{++})
        (64,FORTRAN)
        (45,Python)
        };
\end{axis}
\end{tikzpicture}
}
}
    \caption{Lines of code (LOC) generated by the AI model including comments for all three examples. We used the tool \textit{cloc} to obtain the lines of code. }
    \label{fig:code:lines:of:code}
\end{figure}

Figure~\ref{fig:code:lines:of:code} shows the lines of code (LOC) for the generated codes for all three examples. The tool \textit{cloc}\footnote{\url{https://linux.die.net/man/1/cloc}} was used to obtain the lines of code including comments. Figure~\ref{fig:code:lines:of:code:numerical} shows the lines of code for the numerical integration example. CUDA has the most lines of code. Probably due to the boilerplate for copying from and to device memory and launching the kernels. FORTRAN, and Python have a few less lines compared to \cpp. R has less lines than Python. Matlab had the least lines of code. Figure~\ref{fig:code:lines:of:code:cg} shows the lines of code for the conjugate gradient solver. Again, CUDA has the most lines of code. FORTRAN and \cpp\ are close, and R, Matlab, and Python are on the lower end. 

Figure~\ref{fig:code:lines:of:code:heat:parallel} shows the lines of code for the parallel heat equation solver. Again the CUDA codes have the most lines of code due to the boilerplate code for copying data from and to the device and launching kernels. R, Matlab, \cpp, and Python are close with the lines of code. FORTRAN has the fewest lines of code. Figure~\ref{fig:code:lines:of:code:heat:distributed} shows the lines of code for the distributed heat equation solver using MPI. By adding the MPI boiler plate code, FORTRAN had the most lines of code; second, was \cpp\; and third was Python.  

\begin{table}[tbp]
    \centering
    \caption{Version of all used tools in this study.}
    \begin{tabular}{cccccc}\toprule
    \cpp\ GCC 12 & FORTRAN GCC 12 & Python 3.12 & Matlab R2024a    \\\midrule
    CUDA 12.4.1  &  OpenMPI 4.1.5 & mpi4py 4.0.1 & R 2024a \\\bottomrule
    \end{tabular}
    \label{tab:code:version}
\end{table}

Previously, we looked at the lines of code for the three examples. However, that does not reflect how difficult it was to write this code or how much effort in developer hours would be required. One way to obtain this estimate is to use \textbf{Co}nstructive \textbf{Co}st \textbf{Mo}del (COCOMO)~\citep{5010193,1237981}. We expect that the COCOMO model should give accurate estimates for the numerical integration and the conjugate gradient solver examples. However, the COCOMO Model is designed for serial code execution and does not take parallel or distributed programming into account. One attempt was made to add parallel programming to the COCOMO \textit{II} model~\citep{miller2018applicability}. Since the 90s, however, discussions in the HPC community emerged about having a similar cost model for parallel and distributed applications. However, no model has been proposed as of the time of this writing. Thus, we used the COCOMO model for the parallel and distributed codes as well the serial codes.

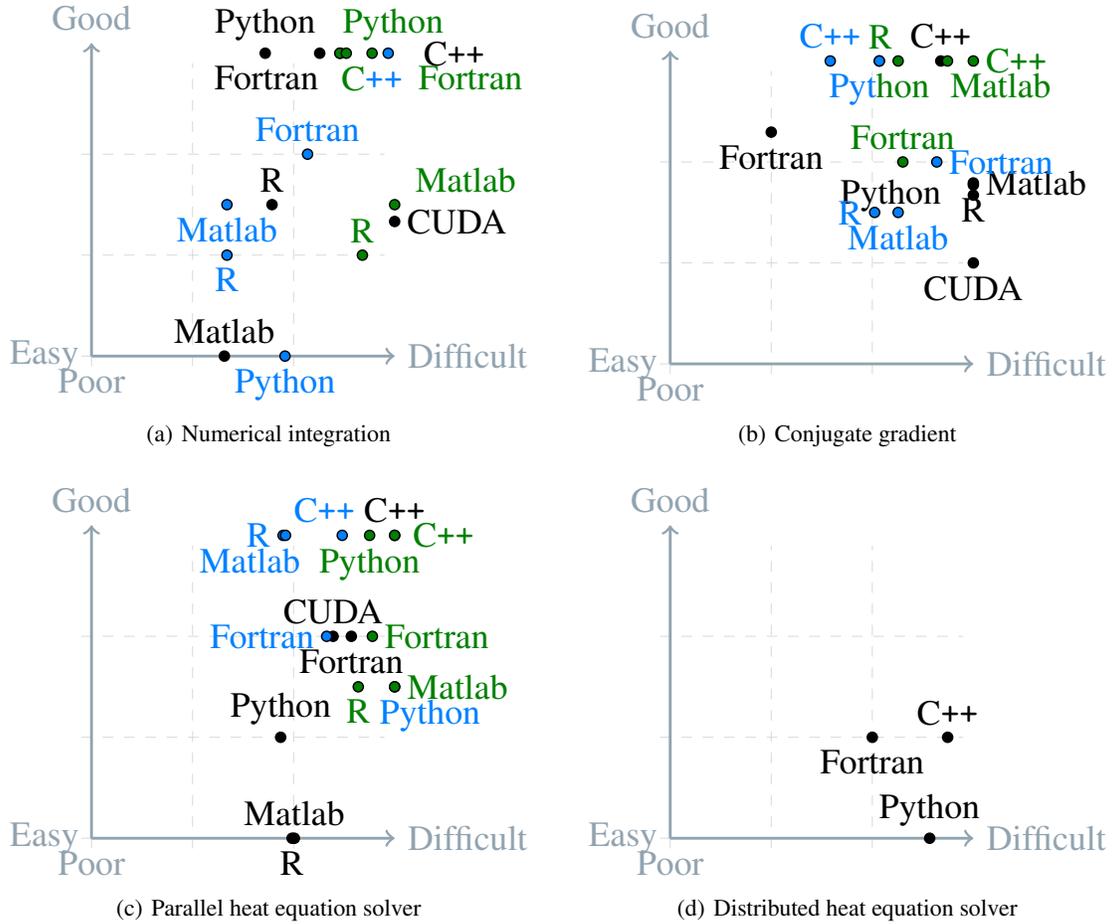
\begin{figure}[tb]
    \centering
    \subfigure[Numerical integration\label{fig:code:quality:scatter:integration}]{
\resizebox{0.45\textwidth}{!}
        {
\begin{tikzpicture}
    \draw[help lines, color=gray!30, dashed] (-0.1,-0.1) grid (2.9,2.9);
    \draw[->,thick,cadetgrey] (0,0)--(3,0) node[right]{Difficult};
    \draw[->,thick,cadetgrey] (0,0)--(0,3.1) node[above,cadetgrey]{Good};
    \node[left,cadetgrey] at (0,0) {Easy};
    \node[below,cadetgrey] at (0,0) {Poor};
    \draw[fill=black] (1.719101124,3) circle [radius=0.05] node[above] {Python}; 
    \draw[fill=black] (2.258426966,3) circle [radius=0.05] node[below] {Fortran\quad \quad \quad \quad};
    \draw[fill=black] (1.314606742,0) circle [radius=0.05] node[above] {Matlab};
    \draw[fill=black] (1.786516854,1.5) circle [radius=0.05] node[above] {R};
    \draw[fill=black] (2.460674157,3) circle [radius=0.05] node[right] {\quad \quad \cpp};
    \draw[fill=black] (3,1.333333334) circle [radius=0.05] node[right] {CUDA};
    \draw[fill=darkgreen] (2.457446809,3) circle [radius=0.05] node[above] {\textcolor{darkgreen}{\quad \quad \quad  Python}};
    \draw[fill=darkgreen] (2.680851064,1) circle [radius=0.05] node[above] {\textcolor{darkgreen}{R}};
    \draw[fill=darkgreen] (2.521276596,3) circle [radius=0.05] node[below] {\textcolor{darkgreen}{\quad \quad \quad \quad \quad \quad \quad Fortran}};
    \draw[fill=darkgreen] (3,1.5) circle [radius=0.05] node[above] {\textcolor{darkgreen}{\quad \quad \quad \quad Matlab}};
    \draw[fill=darkgreen] (2.776595745,3) circle [radius=0.05] node[below] {\textcolor{darkgreen}{C}\textcolor{azure}{\texttt{++}}};
    \draw[fill=azure] (1.914893617,0) circle [radius=0.05] node[below] {\textcolor{azure}{Python}};
    \draw[fill=azure] (1.340425532,1) circle [radius=0.05] node[below] {\textcolor{azure}{R}};
    \draw[fill=azure] (2.138297872,2) circle [radius=0.05] node[above] {\textcolor{azure}{Fortran}};
    \draw[fill=azure] (1.340425532,1.5) circle [radius=0.05] node[below] {\textcolor{azure}{Matlab}};
    \draw[fill=azure] (2.936170213,3) circle [radius=0.05] node[below] {\textcolor{azure}{}};
    \end{tikzpicture}

}
}
    \subfigure[Conjugate gradient\label{fig:code:quality:scatter:cg}]{
\resizebox{0.45\textwidth}{!}
        {
\begin{tikzpicture}
    \draw[help lines, color=gray!30, dashed] (-0.1,-0.1) grid (2.9,2.9);
    \draw[->,thick,cadetgrey] (0,0)--(3,0) node[right]{Difficult};
    \draw[->,thick,cadetgrey] (0,0)--(0,3.1) node[above,cadetgrey]{Good};
    \node[left,cadetgrey] at (0,0) {Easy};
    \node[below,cadetgrey] at (0,0) {Poor};
    \draw[fill=black] (3,1.67114094) circle [radius=0.05] node[left] {Python\;\;}; 
    \draw[fill=black] (1,2.295302013) circle [radius=0.05] node[below] {Fortran};
    \draw[fill=black] (3,1.791946309) circle [radius=0.05] node[right] {Matlab};
    \draw[fill=black] (3,1.771812081) circle [radius=0.05] node[below] {R};
    \draw[fill=black] (2.677852349,3) circle [radius=0.05] node[above] {\cpp};
    \draw[fill=black] (3,1) circle [radius=0.05] node[below] {CUDA};
    \draw[fill=darkgreen] (2.069767442,3) circle [radius=0.05] node[below] {\textcolor{red}{}}; 
    \draw[fill=darkgreen] (3,3) circle [radius=0.05] node[right] {\textcolor{darkgreen}{\cpp}};
    \draw[fill=darkgreen] (2.302325581,2) circle [radius=0.05] node[above] {\textcolor{darkgreen}{Fortran}}; 
    \draw[fill=darkgreen] (2.255813953,3) circle [radius=0.05] node[above] {\textcolor{darkgreen}{R\quad}}; 
    \draw[fill=darkgreen] (2.744186047,3) circle [radius=0.05] node[below] {\textcolor{darkgreen}{\quad \quad \quad Matlab}};
    \draw[fill=azure] (2.069767442,3) circle [radius=0.05] node[below] {\textcolor{azure}{Pyt}\textcolor{darkgreen}{hon}}; 
    \draw[fill=azure] (1.583850932,3) circle [radius=0.05] node[above] {\textcolor{azure}{\cpp}};
    \draw[fill=azure] (2.637931034,2) circle [radius=0.05] node[right] {\textcolor{azure}{Fortran}}; 
    \draw[fill=azure] (2.023255814,1.5) circle [radius=0.05] node[left] {\textcolor{azure}{R}}; 
    \draw[fill=azure] (2.255813953,1.5) circle [radius=0.05] node[below] {\textcolor{azure}{Matlab}};
  
    \end{tikzpicture}
}
}

    \subfigure[Parallel heat equation solver\label{fig:code:quality:scatter:parallel}]{
\resizebox{0.45\textwidth}{!}
        {
\begin{tikzpicture}
    \draw[help lines, color=gray!30, dashed] (-0.1,-0.1) grid (2.9,2.9);
    \draw[->,thick,cadetgrey] (0,0)--(3,0) node[right]{Difficult};
    \draw[->,thick,cadetgrey] (0,0)--(0,3.1) node[above,cadetgrey]{Good};
    \node[left,cadetgrey] at (0,0) {Easy};
    \node[below,cadetgrey] at (0,0) {Poor};
    \draw[fill=black] (1.872180451,1) circle [radius=0.05] node[above] {Python}; 
    \draw[fill=black] (2.571428571,2) circle [radius=0.05] node[below] {Fortran};
    \draw[fill=black] (2.007518797,0) circle [radius=0.05] node[above] {Matlab};
    \draw[fill=black] (1.984962406,0) circle [radius=0.05] node[below] {R};
    \draw[fill=black] (3,3) circle [radius=0.05] node[above] {\cpp};
    \draw[fill=black] (2.390977444,2) circle [radius=0.05] node[above] {CUDA};
    \draw[fill=azure] (3,1.5) circle [radius=0.05] node[below] {\textcolor{azure}{\quad \quad Python}}; 
    \draw[fill=azure] (2.481308411,3) circle [radius=0.05] node[above] {\textcolor{azure}{\cpp\quad}};
    \draw[fill=azure] (2.327102804,2) circle [radius=0.05] node[left] {\textcolor{azure}{Fortran}}; 
    \draw[fill=azure] (1.896,3) circle [radius=0.05] node[left] {\textcolor{azure}{R}}; 
    \draw[fill=azure] (1.92,3) circle [radius=0.05] node[below] {\textcolor{azure}{Matlab\quad\quad}};
    \draw[fill=darkgreen] (2.752293578,3) circle [radius=0.05] node[below] {\textcolor{darkgreen}{Python}}; 
    \draw[fill=darkgreen] (3,3) circle [radius=0.05] node[right] {\textcolor{darkgreen}{\,\cpp}};
    \draw[fill=darkgreen] (2.779816514,2) circle [radius=0.05] node[right] {\textcolor{darkgreen}{Fortran}}; 
    \draw[fill=darkgreen] (2.64,1.5) circle [radius=0.05] node[below] {\textcolor{darkgreen}{R}}; 
    \draw[fill=darkgreen] (3,1.5) circle [radius=0.05] node[right] {\textcolor{darkgreen}{Matlab}}; 
    \end{tikzpicture}
}
}
    \subfigure[Distributed heat equation solver\label{fig:code:quality:scatter:distributed}]{
\resizebox{0.45\textwidth}{!}
        {
\begin{tikzpicture}
    \draw[help lines, color=gray!30, dashed] (-0.1,-0.1) grid (2.9,2.9);
    \draw[->,thick,cadetgrey] (0,0)--(3,0) node[right]{Difficult};
    \draw[->,thick,cadetgrey] (0,0)--(0,3.1) node[above,cadetgrey]{Good};
    \node[left,cadetgrey] at (0,0) {Easy};
    \node[below,cadetgrey] at (0,0) {Poor};
    \draw[fill=black] (2.56779661,0) circle [radius=0.05] node[above] {Python}; 
    \draw[fill=black] (2,1) circle [radius=0.05] node[below] {Fortran};
    \draw[fill=black] (2.745762712,1) circle [radius=0.05] node[above] {\cpp};
    \end{tikzpicture}
}
}
    \caption{Classification of the effort to implement the generated codes from \textbf{easy} to \textbf{difficult} using the Constructive Cost Model (COCOMO). We also classify the quality of the generated code from \textbf{poor} to \textbf{good} using the attributes: compilation, runtime, and correctness. For comparison with other models, we added results for ChatGPT in \textcolor{azure}{blue} and DeepSeek \textcolor{darkgreen}{green} from the authors previous work \cite{diehl2024evaluating} and \cite{nader2025llmhpcbenchmarkingdeepseeks}, respectively. Note that for both the \cpp\ and Python codes, the values for ChatGPT and DeepSeek for the conjugate gradient were so close that the label is colored in green and blue. }
    \label{fig:code:quality:scatter}
\end{figure}

In particular, we use the following metric to classify the code quality from \textbf{poor} to \textbf{good}
\begin{align}
    q(language) := \left( \frac{computation + runtime + correctness}{n} \right)
\end{align}
where $n=3$ for \cpp, Fortran, and CUDA; $n=2$ for Python, Matlab, and R; and $computation,runtime,correctness\in {0,1}$. We use \textit{scc}\footnote{\url{https://github.com/boyter/scc}} to get the COCOMO estimates for the number of person months it would take a human to write the code. We use the estimated person months to classify whether it was \textbf{easy} or \textbf{difficult} to develop the code. Figure~\ref{fig:code:quality:scatter} shows the results.





\section{Performance of the generated codes}
\label{sec:performance:generated}
\begin{figure}[tb]
    \centering
     \subfigure[\label{fig:generated:scaling:cg}]{
\resizebox{0.45\textwidth}{!}
        {
\begin{tikzpicture}
\begin{axis}[ylabel=average time (s),grid,legend,title=Conjugate gradient,ymode=log,log basis y={2},xmode=log,
       log basis x={2},legend pos=north west,xlabel=matrix size $(N \times N)$]
\addplot[black,mark=square*] table [x=matrix_size (NxN), y=avg(s), col sep=comma] {generation_cg_cpp.csv};
\addplot[black,mark=diamond*] table [x=matrix_size (NxN), y=avg(s), col sep=comma] {generation_cg_fortran.csv};
\addplot[black,mark=*] table [x=matrix_size (NxN), y=avg(s), col sep=comma] {generation_cg_python.csv};
\legend{\cpp,Fortran,Python}
\end{axis}
\end{tikzpicture}
}
}
 \subfigure[\label{fig:generated:scaling:heat}]{
\resizebox{0.45\textwidth}{!}
        {
\begin{tikzpicture}
\begin{axis}[title=Parallel heat equation solver,xlabel=\# cores,ylabel=average time (s),grid,ymode=log,log basis y={2},xmode=log,log basis x={2},legend,legend style={at={(0.8,0.7)}}]
\addplot[black,mark=square*] table [x=omp_threads, y=avg(s), col sep=comma] {omp_heat_shared_cpp.csv};
\addplot[black,mark=diamond*] table [x=omp_threads, y=avg(s), col sep=comma] {omp_heat_shared_fortran.csv};
\addplot[black,mark=*] table [x=omp-threads, y=avg(s), col sep=comma] {omp_heat_shared_fixed_python.csv};
\legend{\cpp,Fortran,Python}
\end{axis}
\end{tikzpicture}   
}
}
    \caption{Performance of the generated codes: \subref{fig:generated:scaling:cg} conjugate gradient with increasing matrix sizes on a single core and \subref{fig:generated:scaling:heat} parallel heat equation solver with increasing number of cores and a fixed problem size of 100,000 nodes and 10,000 time steps. For the Python code the problem size had to be reduced to 1024 nodes since the larger size took too long.}
    \label{fig:generated:scaling}
\end{figure}
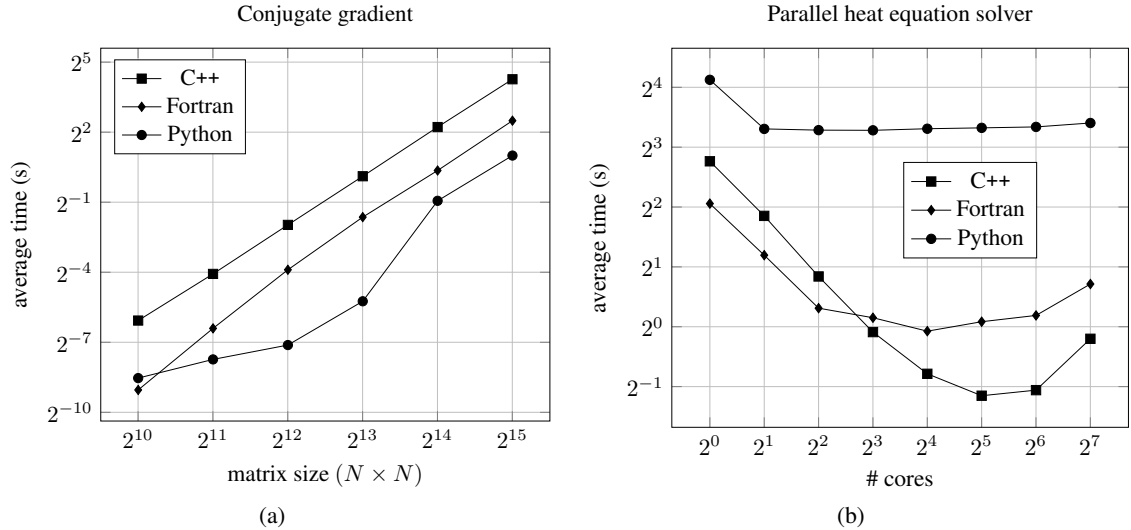

For the generated codes, the performance of the conjugate gradient and the performance of the parallel heat equation solver were analyzed. The distributed heat equation solver and CUDA codes were not considered, since the code quality was so low. Figure~\ref{fig:generated:scaling} shows the results. For all runs, we plot the average run time among five runs. Figure~\ref{fig:generated:scaling:cg} shows the scaling for the conjugate gradient example for increasing matrix sizes $N \times N$ with $N=\{1024,2024,4096,16384,32768\}$. Note the conjugate gradient code is serial, and we can only observe the behavior of the increasing runtime with the matrix size. We had to modify the code to allow for variable matrix sizes and measuring the run time. Figure~\ref{fig:generated:scaling:heat} shows the scaling for the parallel heat equation solver with increasing number of cores (from a single core to 128 cores) and a fixed problem size of 100,000 nodes and 10,000 time steps. For the Python code, the problem size had to be reduced to 1024 nodes since the larger size took too long. The Python code scaled from one core going to two cores. After that, run time increases slightly with increasing cores. The Fortran codes scaled up to three cores and the \cpp\ code scaled up to 32 cores. This behavior is similar to the codes generated with DeepSeek in the author's previous work~\cite{nader2025llmhpcbenchmarkingdeepseeks}. All runs were done using an AMD\ EPYC\ 7742\ 64-Core Processor with 2 sockets and 128 cores in total.

\section{Code documentation}
\label{sec:code:documentation}
We did not include the numerical integration codes in the documentation test since the example was too simple. Note that during the code generation, the LLM added comments to some of the codes. We asked the AI model the following query: 
\begin{displayquote}
\begin{tcolorbox}[colback=gray!15]

Write documentation for the following \textbf{\{Python,\cpp,Fortran\}} code
\end{tcolorbox}
\end{displayquote}
We validated whether all of the function arguments were correct and correctly documented, and measured the ratio of lines of code and lines of documentation.

\subsection{\cpp}
For the conjugate gradient, all function arguments were documented, and 80 lines of code and 30 lines of documentation were generated. For the heat equation solver, no function arguments were in the generated code. The code had 38 lines of code and 50 lines of documentation.
\begin{displayquote}
\begin{tcolorbox}[colback=gray!15]

Write documentation for the following \textbf{\cpp,} code using doxygen
\end{tcolorbox}
\end{displayquote}
Note we had to ask for doxygen comments because without the LLM just described the code's functionality in text form.

\subsection{FORTRAN}
For the conjugate gradient solver, 57 lines of documentation were added to the 50 lines of code. The function arguments were not documented. For the heat equation solver, 26 lines of documentation were added to the 31 lines of code. The generated code had no functions with arguments.

\subsection{Python}
The generated code for the conjugate gradient and heat equation solver already had documentation, including function arguments. The generated code for the conjugate included 14 lines of documentation, and the code for the heat equation solver included 35 comments, respectively.

\subsection{R}
The R code generated for the conjugate gradient had some comments, but when the model was asked to produce documentation, it added valuable comments, \emph{e.g.}\ documentation of the function arguments. The generated code had 25 lines of comments. The generated documentation added not much value, \emph{e.g.}\ no function arguments were documented. 16 lines of documentation were added to the 16 lines of code.

\begin{lstlisting}[language=matlab,caption=Generated documentation for the Matlab heat equation solver. ,label=matlab:documentation:example,escapechar=|, float=false]
% Define the parallel solver function
% Solves the 1D heat equation using the Euler method with parallel processing
function u = solve_heat_equation_parallel(u0, dx, dt, alpha, t_end, num_workers)
    % Get the number of grid points
    nx = length(u0);
    % Compute the number of time steps
    nt = ceil(t_end / dt);
    % Initialize the solution vector
    u = u0;
    
    % Create a parallel pool of workers
    parpool(num_workers);
    
    % Time-stepping loop
    for i = 1:nt
        % Compute the derivative using the heat equation
        du_dt = heat_equation(u, dx, dt, alpha);
        % Update the solution using the Euler method
        u = euler_step(u, du_dt, dt);
    end
    
    % Delete the parallel pool
    delete(gcp('nocreate'));
end
\end{lstlisting}

\subsection{Matlab}
Listing~\ref{matlab:documentation:example}) shows an example of the generated Matlab documentation. For the conjugate gradient solver, the code had 19 lines of documentation added. No function documentation was generated. For the heat equation solver, 19 lines of documentation were added to the 26 lines of generated code. No function arguments were documented. Note we had to add that the LLM should generate inline documentation to the textual description of the code generated. 


\begin{table}[tb]
    \centering
    \rowcolors{2}{gray!25}{white}
    \begin{tabular}{l|ccccc}\toprule
     Language    & \cpp & FORTRAN & Python & Matlab & R  \\\midrule
     Attribute  & \multicolumn{5}{c}{Conjugate gradient} \\ \midrule
     LOC  & 80 & 50 & 27 & 33 & 27 \\
     LOD  & 30 & 57 & \textcolor{azure}{14} & \textcolor{azure}{19} & 25\\
     Ratio & 2.6 & 0.9 & 1.9 & 1.7 & 1.1\\\midrule
     Attribute  & \multicolumn{5}{c}{Heat equation solver} \\ \midrule
     LOC  & 38 & 31 & 41 & 26 & 16\\
     LOD  & 50 & 26 & \textcolor{azure}{35} & 19 & 16\\
     Ratio & 0.8 & 1.2 & 1.2 & 1.3 & 1\\\midrule
    \end{tabular}
    \caption{Information about the generated documentation: lines of of code (LOC) of the generated code and line of documentation added by the LLM. We show the ratio of between the loc and lod. The blue numbers indicate that the generated code already had documentation even we did not ask the LLM to generate code.}
    \label{tab:my_label}
\end{table}

\section{Generating unit tests}
\label{sec:code:tests}
We asked the AI model the following query: 
\begin{displayquote}
\begin{tcolorbox}[colback=gray!15]

Write a unit test for the following \textbf{\{Python,\cpp,Fortran\}} code
\end{tcolorbox}
\end{displayquote}
to generate the unit tests. This was followed by the generated code in the query.

\subsection{\cpp}
For \cpp\, the LLM decided to use \textit{GoogleTest}\footnote{\url{https://github.com/google/googletest}} and generated the code depicted in Listing~\ref{cpp:testing} for the \texttt{riemann\_sum()}. It incorrectly assumed that there was a header file associated with the function to be tested. So we copied the generated \cpp code
 into this header and modified it to have the expected arguments and types. In addition, we integrated the test into CMake to be able to run the \lstinline[language=bash]{ctest} command. We used the Linux tool \lstinline[language=bash]{gcovr}\footnote{\url{https://github.com/gcovr/gcovr}} to estimate the code coverage of the generated tests. Unit tests were generated on 01/06/2024.

%
%

\begin{lstlisting}[language=c++,caption=Generated unit tests for the \cpp numerical integration example.,label=cpp:testing,escapechar=|, float=false]
#include <gtest/gtest.h>
#include "riemann_sum.h" // assume the code is in a header file called riemann_sum.h

TEST(RiemannSumTest, SimpleTest) {
  // Test the Riemann sum for a simple case
  double a = 0.0;
  double b = M_PI / 2.0;
  int n = 1000;
  double expected_result = 1.0; // exact value of the integral
  double tolerance = 1e-3; // tolerance for the comparison

  double result = riemann_sum(a, b, n);
  EXPECT_NEAR(result, expected_result, tolerance);
}

TEST(RiemannSumTest, NegativeLimitsTest) {
  // Test the Riemann sum for a case with negative limits
  double a = -M_PI / 2.0;
  double b = 0.0;
  int n = 1000;
  double expected_result = -1.0; // exact value of the integral
  double tolerance = 1e-3; // tolerance for the comparison

  double result = riemann_sum(a, b, n);
  EXPECT_NEAR(result, expected_result, tolerance);
}

TEST(RiemannSumTest, LargeNTest) {
  // Test the Riemann sum for a large number of intervals
  double a = 0.0;
  double b = M_PI;
  int n = 100000;
  double expected_result = 2.0; // exact value of the integral
  double tolerance = 1e-6; // tolerance for the comparison

  double result = riemann_sum(a, b, n);
  EXPECT_NEAR(result, expected_result, tolerance);
}

int main(int argc, char **argv) {
  ::testing::InitGoogleTest(&argc, argv);
  return RUN_ALL_TESTS();
}

\end{lstlisting}

\newpara
\noindent \textbf{Numerical integration} Here, the LLM generated three different tests namely: \lstinline[language=bash]{SimpleTest}, \lstinline[language=bash]{NegativeLimitsTest}, and \lstinline[language=bash]{LargeNTest}. The \lstinline[language=bash]{SimpleTest} evaluates the integral from zero to $\pi/2$ for $1000$ iterations. And compares the computed result with the exact solution which is $-1$ for a tolerance of $10^{-3}$. The \lstinline[language=bash]{NegativeLimitsTest} computes the integral from $\pi/2$ to zero. The \lstinline[language=bash]{LargeNTest} repeats the first test using $100000$ iterations and a tolerance of $10^{-6}$. The code coverage was $100$\%.

 \newpara
\noindent \textbf{Conjugate gradient} Here, the LLM generated two test cases namely \lstinline[language=bash]{SimpleTest} and \lstinline[language=bash]{LargeMatrixTest}. The \lstinline[language=bash]{SimpleTest} failed due to a wrong solution $x=(0.5\quad 1.0\quad 1.5)$ for the proposed matrix system 
    \begin{align*}
        A = \left( \begin{matrix}
            4 &  -1 & 0 \\ -1 &  4 &  -1 \\ 0 & -1 & 4
        \end{matrix} \right)  \text{ and }  b = \left( \begin{matrix} 1 \\ 2 \\ 3  \end{matrix} \right) \text{.}
    \end{align*}
    After updating the solution to $x=(0.464 0.857 0.964)$ which was validated with numpy package the first test passes. For the second test the matrix system
    \begin{align*}
        A^{n \times n} = \left( \begin{matrix}
             4 & -1 & 0 & 0 & 0 & 0\\
             0 & -1 & 4  & -1 & 0 & 0\\
             0 & -1 & 4 & -1 & 0 & 0\\
             0 & 0 & \ddots & \ddots & \ddots & 0 \\
             0 & 0 & -1 & 4 & -1 & 0  \\
             0 & 0 & 0 & -1 & 4 & -1  \\
             0 & 0 & 0 &  0 & -1 & 4 
        \end{matrix} \right) \text{ and }  b = \left( \begin{matrix} 1 \\ 2 \\ 3 \\ \vdots  \\ n-2 \\ n-1 \\ n \end{matrix} \right) 
    \end{align*}
    was proposed to be solved for $n=10$. If a ten by ten matrix is large is a different topic to be discussed. However, the solution $x^n=1/4 (1,2,\ldots,n-1,n)$ was wrong. We computed the solution for ten elements using numpy and validated against this solutions and the test worked. The test coverage was $100$\%.

\newpara
 \noindent\textbf{Heat equation solver}: The generated unit test did not compile since the variables: \lstinline[language=c++]{NT}, \lstinline[language=c++]{NX}, \lstinline[language=c++]{NT}, \lstinline[language=c++]{DX}, \lstinline[language=c++]{DT}, and \lstinline[language=c++]{ALPHA} were not defined. In addition functions were declared in the wrong order and the last two functions to generate the initial displacement and compute a single step of the heat equation had to be moved to the top. Three tests were generated \lstinline[language=bash]{InitialCondition}, \lstinline[language=bash]{SingleTimeStep}, and \lstinline[language=bash]{MultipleTimeSteps}. The displacement $u$ is initialized as
    \begin{align*}
         u_i = \sin(\pi \cdot i \cdot DX),\quad i=0,1,\ldots,NX-2,NX-1\text{.}
    \end{align*}
    The first test checks if the displacement is correctly initialized. The \lstinline[language=bash]{SingleTimeStep} and \lstinline[language=bash]{MultipleTimeSteps} compared if the single threaded implementation and the multi threaded implementation compute the same result. The LLM mentioned that a better test would be to compare against a analytical solution. The test coverage was 100\%.

\subsection{FORTRAN}
The tests were generated 01/27/2015. As for the \cpp\ unit tests the Google test framework was used. However, the unit test was written in FORTRAN. First, the generated code did not compile. After, fixing the function call \lstinline[language=fortran]{assert_equals} was not defined. After some research, the Google Test framework is not available in FORTRAN. The way to go would be to extract the generated test cases and use the module \lstinline[language=fortran]{iso_fortran_env} to provide binding to \cpp. After that the Google test related framework specific function will be called within the \cpp\ program and the unit tests written in FORTRAN are called. The LLM did not generate these binding and provided the \cpp\ code. So we would have to do all the work. Therefore, we decided not to write the unit test ourself.

\subsection{Python}
The unit tests were generated 01/25/2025. The AI model used the Python package \lstinline[language=python]{unittest}\footnote{\url{https://docs.python.org/3/library/unittest.html}}.

\newpara
\noindent \textbf{Numerical integration:} The generated unit test used the function \lstinline[language=python]{f(x)} defined in the code was not added to the unit test. After adding the function definition, the code executed. One test was generated to test if the function \lstinline[language=python]{f(x)} returns the same value as  \lstinline[language=python]{np.sin(x)}. The second test checks if the computed some matches some analytical value. However, the analytical value was wrong and the test failed. The code coverage was 100\%.

 \newpara
\noindent \textbf{Conjugate gradient:} In total, four test cases were written. For the first three test cases, the equation system
    \begin{align*}
        A = \left (\begin{matrix}
            4 & -1 & 0 \\ -1 &  4 & -1 \\ 0 &  -1 & 4
        \end{matrix} \right)
        \text{ and } b = \left(\begin{matrix}
            1 \\ 2 \\3
        \end{matrix} \right) \text{.}
    \end{align*}
    In the first test solves the system and compares the result of \lstinline[language=python]{np.linalg.solve} with a tolerance of $\epsilon=10^{-5}$ and a tolerance of $\epsilon=10^{-8}$, respectively. The third tests used 50 maximal iterations instead of 100 maximal iterations and a tolerance of $\epsilon=10^{-5}$. The last tests checks if the solver fails to solve the singular matrix 
    \begin{align*}
        A = \left(\begin{matrix}
            1 & 1 \\ 
            1 & 1 
        \end{matrix} \right) \text{ and } b= \left(\begin{matrix}
            1 \\ 1
        \end{matrix} \right) \text{.}
    \end{align*}
    The code coverage was 100\%.

 \newpara
\noindent \textbf{Heat equation solver:} The same initial conditions and analytical solution described in Section~\ref{sec:model:problems:heat} was used. The heat equation was solved for one thread, two threads, and for four threads. The following parameters were used:  $L= 1.0$, $n=100$, $h=0.01$, $T=0.1$, $dt=0.1/1000$, and $\alpha=0.1$. The code coverage was 100\%.

\section{Quality of the generated unit tests}
\label{sec:tests:quality}
Table~\ref{tab:unit:tests:eval} shows the results for the generated unit tests.
For \cpp, three unit tests were generated for both the numerical integration and the heat equation solver. All unit tests needed to be corrected. For the heat equation solver, the unit tests did not even compile.

For Python, for the numerical integration problem, two unit tests were generated, but only one was correct. For the conjugate gradient, four correct unit tests were generated, and for the heat equation solver, three correct unit tests were generated.

The model could not generated unit tests for FORTRAN.


\begin{table}[tb]
    \centering
    \rowcolors{2}{gray!25}{white}
    \begin{tabular}{l|ccc|ccc|ccc} 
    
     Language & \rotatebox{60}{Python} & \rotatebox{60}{FORTRAN} & \rotatebox{60}{\cpp} & \rotatebox{60}{Python} & \rotatebox{60}{FORTRAN} & \rotatebox{60}{\cpp} & \rotatebox{60}{Python} & \rotatebox{60}{FORTRAN} & \rotatebox{60}{\cpp} \\\midrule

    Attribute & \multicolumn{3}{c|}{Numerical Integration} & \multicolumn{3}{c|}{Conjugate Gradient} & \multicolumn{3}{c}{Heat Equation Solver} \\\midrule
    Compilation    & -- & NA & \checkmark & -- & NA & \checkmark & -- & NA & \tikzxmark \\
    Test cases     & 2/1 & NA & 3/3 & 4/4 & NA & 2/0 & 3/3 & NA & 3/3 \\
    Execution & \tikzxmark & NA & \checkmark & \checkmark & NA & \checkmark & \checkmark & NA & \checkmark \\
    \bottomrule
    \end{tabular}
     \caption{Summary of the generated test cases for Python, FORTRAN, and \cpp. We checked whether the generated code for the test compiled and executed. We reported the number of generated test cases (first number) and the number of working test cases (second number).}
    \label{tab:unit:tests:eval}
\end{table}

\begin{figure}[tb]
    \centering
    \subfigure[Numerical integration]{
    \begin{tikzpicture}
    \draw[help lines, color=gray!30, dashed] (-0.1,-0.1) grid (2.9,2.9);
    \draw[->,thick,cadetgrey] (0,0)--(3,0) node[right]{Difficult};
    \draw[->,thick,cadetgrey] (0,0)--(0,3.1) node[above,cadetgrey]{Good};
    \node[left,cadetgrey] at (0,0) {Easy};
    \node[below,cadetgrey] at (0,0) {Poor};
    \draw[->,thick,cadetgrey] (0,0)--(3,0) node[right]{Difficult};
    \draw[->,thick,cadetgrey] (0,0)--(0,3.1) node[above,cadetgrey]{Good};
    \draw[fill=gray] (2.52,0.75) circle [radius=0.05] node[left] {Python}; 
    \draw[fill=gray] (3,3) circle [radius=0.05] node[above] {\cpp};
    \end{tikzpicture}
    }
    \subfigure[Conjugate gradient]{
    \begin{tikzpicture}
    \draw[help lines, color=gray!30, dashed] (-0.1,-0.1) grid (2.9,2.9);
    \draw[->,thick,cadetgrey] (0,0)--(3,0) node[right]{Difficult};
    \draw[->,thick,cadetgrey] (0,0)--(0,3.1) node[above,cadetgrey]{Good};
    \draw[->,thick,cadetgrey] (0,0)--(3,0) node[right]{Difficult};
    \draw[->,thick,cadetgrey] (0,0)--(0,3.1) node[above,cadetgrey]{Good};
    \node[left,cadetgrey] at (0,0) {Easy};
    \node[below,cadetgrey] at (0,0) {Poor};
    \draw[fill=black] (3,3) circle [radius=0.05] node[left] {Python}; 
    \draw[fill=black] (2.58,2) circle [radius=0.05] node[above] {\cpp}; 
    \end{tikzpicture}
    }
    \subfigure[Heat equation solver]{
    \begin{tikzpicture}
    \draw[help lines, color=gray!30, dashed] (-0.1,-0.1) grid (2.9,2.9);
    \draw[->,thick,cadetgrey] (0,0)--(3,0) node[right]{Difficult};
    \draw[->,thick,cadetgrey] (0,0)--(0,3.1) node[above,cadetgrey]{Good};
    \draw[help lines, color=gray!30, dashed] (-0.1,-0.1) grid (2.9,2.9);
    \draw[->,thick,cadetgrey] (0,0)--(3,0) node[right]{Difficult};
    \draw[->,thick,cadetgrey] (0,0)--(0,3.1) node[above,cadetgrey]{Good};
    \node[left,cadetgrey] at (0,0) {Easy};
    \node[below,cadetgrey] at (0,0) {Poor};
    \draw[fill=azure] (3,3) circle [radius=0.05] node[left] {Python}; 
    \draw[fill=azure] (2.8,2) circle [radius=0.05] node[below] {\cpp};
    \end{tikzpicture}
    }
    \caption{Classification of the effort to implement the generated unit tests by a developer from \textbf{easy} to \textbf{difficult} using the Constructive Cost Model (COCOMO). The quality of the translated code was classified from \textbf{poor} to \textbf{good} was computed using the attributes: compilation, runtime, and correctness.}
    \label{fig:unit:test:quality:scatter}
\end{figure}
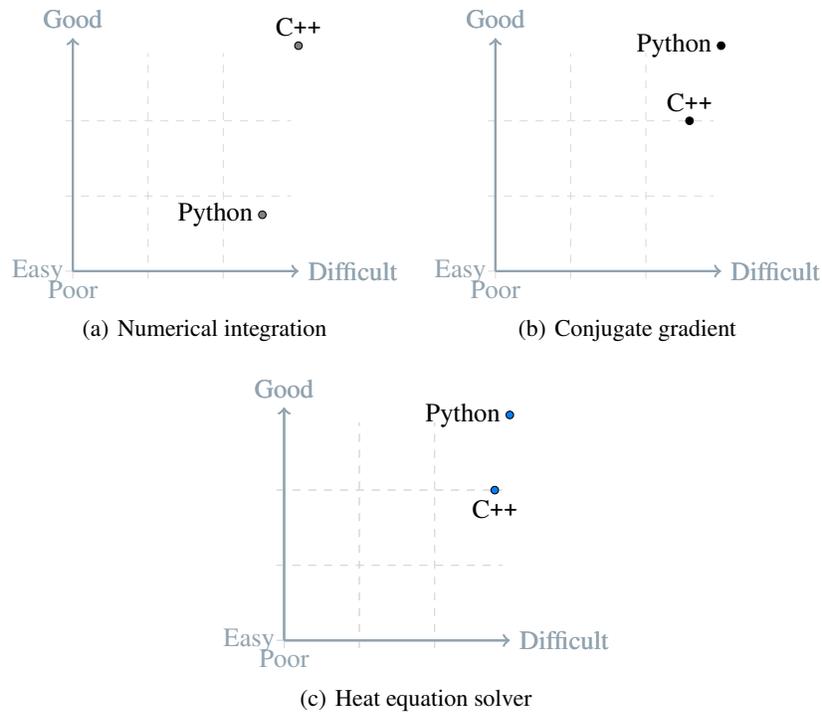

For all unit tests generated, the code coverage was 100\% after fixing the errors in the unit tests. Figure~\ref{fig:unit:test:quality:scatter} shows the classification for the generated unit tests. For the numerical integration problem, \cpp\ performed better. For the two other applications, Python performed better. For all three codes, the effort to implement the unit tests was comparable. However, different numbers of unit test were generated for both languages.

\section{Translation code between programming languages }
\label{sec:code:translated}

\subsection{FORTRAN to Python and \cpp}
For the translation of the code, we used the previously generated FORTRAN examples. However, in all cases, the results needed to be corrected. We provide the following query to the AI model
\begin{displayquote}
\begin{tcolorbox}[colback=gray!15]

Translate the FORTRAN code below to \textbf{\{\cpp, Python\}}.
\end{tcolorbox}
\end{displayquote}
The numerical integration example, the conjugate gradient solver, and parallel heat equation solver were translated by the AI model on 10/01/2024. We used the tool \textit{wdiff -s}\footnote{\url{https://www.commandlinux.com/man-page/man1/wdiff.1.html}} to obtain the differences between the translated Python code and the generated Python code.

\subsubsection{\cpp} \label{translated_code:fortran_to_cpp_mpi}
 
\textbf{Numerical Integration.}
The numerical integration code was compiled, executed and provided correct results. Compared to the previously generated C++ code, 89\% of the lines of code were the same, one percent was deleted, and 10\% were changed.

\newpara
\noindent
\textbf{Conjugate gradient.}
The translated code did not compile due to a missing closing parenthesis for the main function and a closing quote for a string. The code executed and computed the correct solution. However, the validation method was not translated. Here, 39\% of the lines of code were the same, 26\% were deleted, and 53\% were changed. 

\newpara
\noindent
\textbf{Heat equation solver.} The translated code did not compile because the code tried to redefine \lstinline[language=c++]{const double M_PI}, which \lstinline[language=c++]{#include <cmath>} defines as a macro. After commenting out the line, the code compiled. In addition, the header \lstinline{#include <omp.h>} was included, but no OpenMP functions were used. The code executed and produced the correct result. Here, 76\% of the lines of code were the same, one percent was deleted, and 23\% changed.

\newpara
\noindent 
\textbf{Distributed heat equation solver} The translated MPI code for the C++ distributed heat equation solver did not compile due to multiple errors in function definitions and in calling MPI routines (passing too many parameters to \lstinline[language=c++]{MPI_Bcast()} and  \lstinline[language=c++]{MPI_Finalize()}). After fixing the compilation errors and the parameters for the heat equation solver, the code compiled, executed and produced the correct result. However, as was the case with the MPI generated codes, the translated code is equivalent to a serial program where each individual processes computed the entire domain. Here, 54\% of the lines of code were the same, 15\% deleted, and 31\% changed.

\subsubsection{Python}

\textbf{Numerical Integration.}
The translated Python code executed and produced the correct results. Here 62\% of the lines of code were common, 26\% were deleted, and 12\% were changed.

\newpara
\noindent 
\textbf{Conjugate gradient.} The translated code had runtime errors. First, the initialization of the residual was wrong because it was initialized as a vector and not a floating point number. With that change, the conjugate gradient solver worked. However, the Gaussian elimination implementation step had issues because the matrix and vector were initialized as integer values. Compared to the previously generated Python code, the difference was that 39\% of the original lines of code were the same, 29\% were deleted, and 32\% changed.

\newpara
\noindent\textbf{Heat equation solver.} The translated code for the parallel heat equation solver worked. However, a different framework, namely numba, was used for parallelism. Here, 30\% were common, one percent was deleted, and 68\% were changed.

\newpara
\noindent\textbf{Distributed heat equation solver} The translated MPI code for the python distributed heat equation solver was almost identical to the \cpp\ translated code (described in subsection \ref{translated_code:fortran_to_cpp_mpi}). After fixing the same errors and the solver parameters, the code executed and produced the correct result. Again, the code is equivalent to a serial program. The MPI communication library is only invoked once, for broadcasting the initial array. Compared to the generated distributed code, 19\% of the lines of code were the same, 39\% were deleted, and 42\% changed. 

\subsection{MatLab to Python and R}
Here, we used the previously generated Matlab examples. Note that, in order to avoid compounding errors, we used the hand-corrected Matlab code as input. We provided to the AI model the following query
\begin{displayquote}
\begin{tcolorbox}[colback=gray!15]

Translate the Matlab code below to \textbf{\{Python, R\}}
\end{tcolorbox}
\end{displayquote}
The numerical integration example, the conjugate gradient solver, and parallel heat equation solver were all translated by the AI model on 12/06/2024.

\subsubsection{Python}

\textbf{Numerical Integration} The translated Python code had runtime errors due to an undefined variable x. After correction, the code executed and produced correct results. Compared to the generated code, 47\% of the lines of code were the same, 5\% were inserted, and 49\% were changed.

\newpara
\noindent \textbf{Conjugate Gradient} The translated Python code had runtime errors due to an incompatible data type (UFuncTypeError). After fixing the error, the code executed and produced desired results. For this example, 39\% of the lines of code were the same, 16\% inserted, and 45\% changed.
    
\newpara
\noindent \textbf{Heat Equation} The translated Python code encountered runtime errors due to an undefined variable \lstinline[language=python]{dt} and the deprecated \lstinline[language=python]{plt.hold(True)} function in Matplotlib. After defining the variable and removing the \lstinline[language=python]{plt.hold(True)} function, the code executed successfully and produced the correct result.
Compared to the previously generated Python code, 65\%  of the lines of code were the same, one percent was inserted, and 34\% were changed. 

\subsubsection{R}
 \textbf{Numerical Integration: } Listing~\ref{matlab:R:translation} presents the translated R code. It encountered an issue due to improper definition of the integrand function \lstinline[language=R]{plt.hold(sin(x))}for the \lstinline[language=R]{integrate()} function. After fixing this issue, the code executed successfully and produced the correct result. Here, 69\% of the lines of code were the same, three percent were inserted, and 29\% were changed.
%
%
\begin{lstlisting}[language=R,caption=Translated matlab code to R for the numerical integration example.,label=matlab:R:translation,escapechar=|, float=false]
# Define the limits of integration
a <- -pi
b <- 2/3 * pi

# Compute the area using the definite integral
area <- integrate(sin(x), a, b)

# Print the result
cat("Area:", area, "\n")

\end{lstlisting}
    
\newpara
\noindent \textbf{Conjugate Gradient: } The translated R code executed successfully and produced the correct results. Here, 61\% of the lines of code were the same, 13\% were inserted, and 25\% were changed.

\newpara
\noindent \textbf{Heat Equation: } The R code encountered an error due to the use of the \lstinline[language=R]{parpool} function, which is not available in R. We replaced it with the appropriate \lstinline[language=R]{makeCluster} and \lstinline[language=R]{parSapply} functions from the parallel package. However, the code still encountered issues, and after fixing the errors, the code executed successfully but did not produce the correct results.  Here, 47\% of the lines of code were the same, 25\% were inserted, and 29\% were changed.

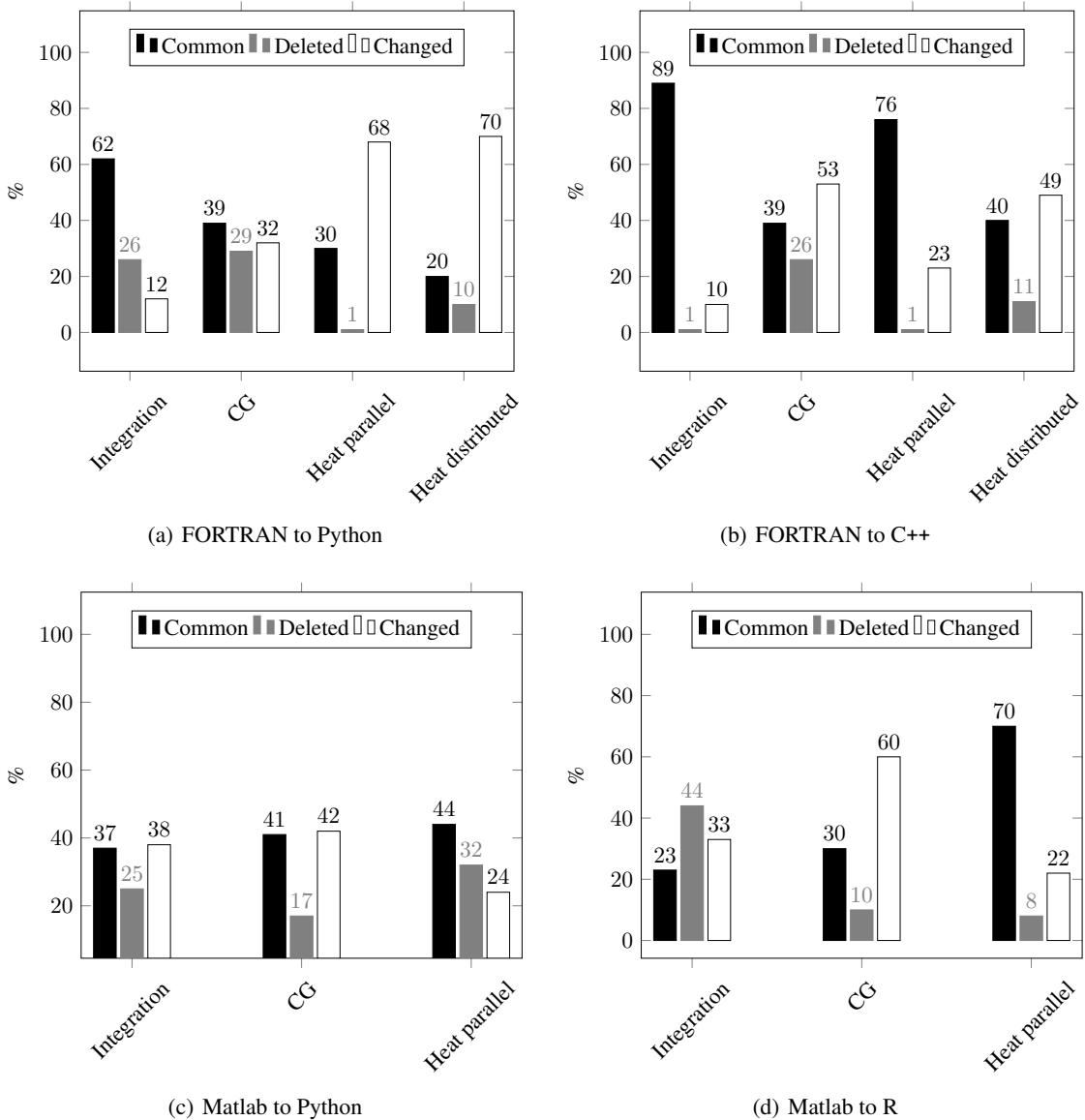
\begin{figure}[tbp]
    \subfigure[FORTRAN to Python\label{fig:translation:difference:python}]{
\resizebox{0.45\textwidth}{!}
        {
        \begin{tikzpicture}
\begin{axis}[
    ybar,
    enlargelimits=0.15,
    legend style={at={(0.5,0.95)},
    anchor=north,legend columns=-1},
    ylabel={\%},
    symbolic x coords={Integration,CG,Heat parallel,Heat distributed},
    xtick=data,
    ymax=100,
    nodes near coords,
    nodes near coords align={vertical},
    xticklabel style={
        rotate=45,
    }
    ]
\addplot[black,fill=black] coordinates {(Integration,62) (CG,39) (Heat parallel,30) (Heat distributed,20)};
\addplot[gray,fill=gray] coordinates {(Integration,26) (CG,29) (Heat parallel,1) (Heat distributed,10)};
\addplot[black,fill=white] coordinates {(Integration,12) (CG,32) (Heat parallel,68) (Heat distributed,70)};
\legend{Common,Deleted,Changed}
\end{axis}
\end{tikzpicture}
}
}
    \subfigure[FORTRAN to \cpp\label{fig:translation:difference:cpp}]{
\resizebox{0.45\textwidth}{!}
        {
        \begin{tikzpicture}
\begin{axis}[
    ybar,
    enlargelimits=0.15,
    legend style={at={(0.5,0.95)},
    anchor=north,legend columns=-1},
    ylabel={\%},
    symbolic x coords={Integration,CG,Heat parallel,Heat distributed},
    xtick=data,
    ymax=100,
    nodes near coords,
    nodes near coords align={vertical},
    xticklabel style={
        rotate=45,
    }
    ]
\addplot[black,fill=black] coordinates {(Integration,89) (CG,39) (Heat parallel,76) (Heat distributed,40)};
\addplot[gray,fill=gray] coordinates {(Integration,1) (CG,26) (Heat parallel,1) (Heat distributed,11)};
\addplot[black,fill=white] coordinates {(Integration,10) (CG,53) (Heat parallel,23) (Heat distributed,49)};
\legend{Common,Deleted,Changed}
\end{axis}
\end{tikzpicture}
}
}

\vspace{0.1cm}

 \subfigure[Matlab to Python\label{fig:translation:difference:matlab:python}]{
\resizebox{0.45\textwidth}{!}
        {
        \begin{tikzpicture}
\begin{axis}[
    ybar,
    enlargelimits=0.15,
    legend style={at={(0.5,0.95)},
    anchor=north,legend columns=-1},
    ylabel={\%},
    symbolic x coords={Integration,CG,Heat parallel},
    xtick=data,
    ymax=100,
    nodes near coords,
    nodes near coords align={vertical},
    xticklabel style={
        rotate=45,
    }
    ]
\addplot[black,fill=black] coordinates {(Integration,37) (CG,41) (Heat parallel,44) };
\addplot[gray,fill=gray] coordinates {(Integration,25) (CG,17) (Heat parallel,32) };
\addplot[black,fill=white] coordinates {(Integration,38) (CG,42) (Heat parallel,24) };
\legend{Common,Deleted,Changed}
\end{axis}
\end{tikzpicture}
}
}
    \subfigure[Matlab to R\label{fig:translation:difference:matlab:r}]{
\resizebox{0.45\textwidth}{!}
        {
        \begin{tikzpicture}
\begin{axis}[
    ybar,
    enlargelimits=0.15,
    legend style={at={(0.5,0.95)},
    anchor=north,legend columns=-1},
    ylabel={\%},
    symbolic x coords={Integration,CG,Heat parallel},
    xtick=data,
    ymax=100,
    nodes near coords,
    nodes near coords align={vertical},
    xticklabel style={
        rotate=45,
    }
    ]
\addplot[black,fill=black] coordinates {(Integration,23) (CG,30) (Heat parallel,70) };
\addplot[gray,fill=gray] coordinates {(Integration,44) (CG,10) (Heat parallel,8) };
\addplot[black,fill=white] coordinates {(Integration,33) (CG,60) (Heat parallel,22) };
\legend{Common,Deleted,Changed}
\end{axis}
\end{tikzpicture}
}
}
    
    \caption{Code similarity for generated and translated codes. In (a) and (b) we translation from Fortran, and in (c) and (d) we see translation from Matlab.}
    \label{fig:translation:difference}
\end{figure}

\section{Quality of the translated code}
\label{sec:translated:quality}
\subsection{FORTRAN to \cpp\ and Python}
Table~\ref{tab:translation:results}(A) shows that for the numerical integration problem, the \cpp\ and Python code both worked without errors. For the conjugate gradient solver, the translated \cpp\ code had compile time issues and the Python code used the wrong types in the Gaussian elimination step.

\begin{table}[tbp]
    \centering
    \begin{tabular}{l|cc|cc|cccc}
      &\multicolumn{6}{c}{\textbf{A. FORTRAN to \cpp  ~and Python}}   \\
        \cmidrule(lr){2-7}  
    & \multicolumn{2}{c|}{Integration} & \multicolumn{2}{c|}{Conjugate gradient} & \multicolumn{2}{c}{Heat equation solver} \\\midrule
    
    Language   & \cpp &  Python & \cpp & Python &  \cpp & Python  \\\midrule 
    \rowcolor{gray!25}   Compilation  & \checkmark  & --   & \tikzxmark & -- & \tikzxmark/\tikzxmark  & --/--  \\
     Runtime  & \checkmark  & \checkmark & \checkmark & \tikzxmark & \checkmark/\tikzxmark & \checkmark/\tikzxmark \\
     \rowcolor{gray!25}  Correctness & \checkmark &  \checkmark & \checkmark & \checkmark &  \checkmark/\tikzxmark  &  \checkmark/\tikzxmark \\\bottomrule

    \addlinespace
    \addlinespace
    &\multicolumn{6}{c}{\textbf{B. Matlab to Python and R}}   \\
    \cmidrule(lr){2-7} 
     & \multicolumn{2}{c|}{Integration} & \multicolumn{2}{c|}{Conjugate gradient} & \multicolumn{2}{c}{Heat equation solver} \\\midrule
    Language   &   Python & R & Python & R & Python & R \\ \midrule 
     Runtime  & \tikzxmark  & \tikzxmark & \tikzxmark & \checkmark & \tikzxmark & \tikzxmark\\
     \rowcolor{gray!25}  Correctness & \checkmark &  \checkmark & \checkmark & \checkmark & \checkmark   & \tikzxmark  \\\bottomrule
    \end{tabular}
   \caption{Results for the translated \textbf{(A)}  FORTRAN codes to \cpp\ or Python, and \textbf{(B)} Matlab codes to Python or R, respectively. We check to see whether the codes compile with the compilers shown in Table~\ref{tab:code:version}. For all codes, we check whether the codes executed without any run time errors, \emph{e.g.}\ index out of bound exceptions. Finally, we checked to see whether the code produces the correct results for the test cases in Section~\ref{sec:model:problems}.}
    \label{tab:translation:results}
\end{table}

Figures~\ref{fig:translation:difference} (a-b) show the difference between the generated \cpp\ or Python code and the generated translation of FORTRAN code to \cpp\ or Python, respectively. We use the Linux tool \textit{wdiff} to evaluate the differences. For the translated Python codes, see Figure~\ref{fig:translation:difference:python}.

For the conversion from FORTRAN to Python, the fewest changes between generated and translated code were observed for numerical integration. For the conjugate gradient solver, the number of changes for these two problems were more balanced. For the parallel heat equation solver, many lines of the Python code were changed since the translated code used a different library for parallel execution.

Figure~\ref{fig:translation:difference:cpp} shows the analysis for the translated code from FORTRAN to \cpp. For the numerical integration, only one line was deleted and 10 lines were changed. For the conjugate gradient it is more balanced with 39 lines in common, 26 lines deleted, and 53 lines changed. For the parallel heat equation, only one line was deleted and 23 lines were changed. For the distributed heat equation solver, 40 lines were common, 11 lines were deleted, and 49 lines were changed.

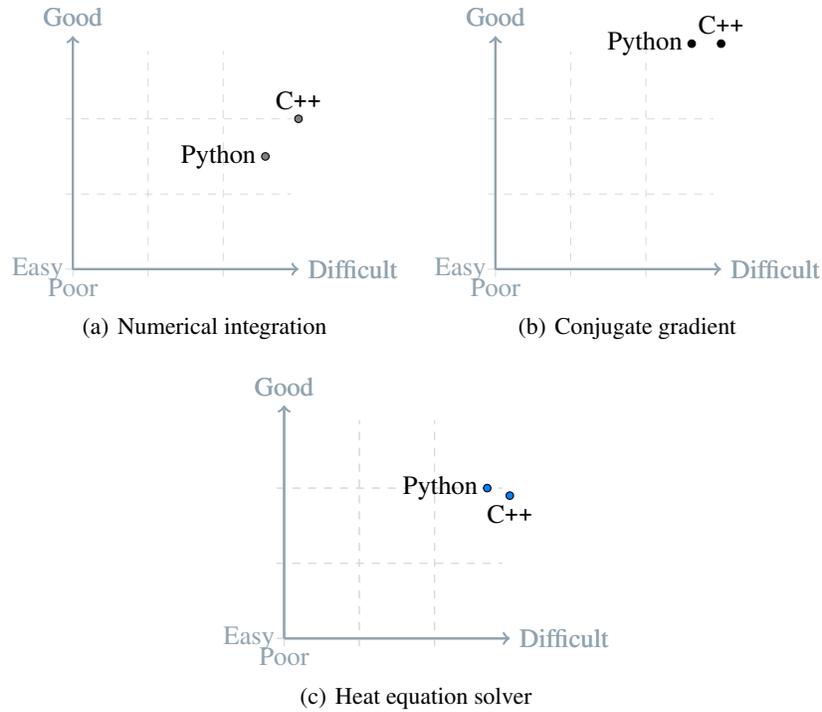
\begin{figure}[tb]
    \centering
    \subfigure[Numerical integration]{
    \begin{tikzpicture}
    \draw[help lines, color=gray!30, dashed] (-0.1,-0.1) grid (2.9,2.9);
    \draw[->,thick,cadetgrey] (0,0)--(3,0) node[right]{Difficult};
    \draw[->,thick,cadetgrey] (0,0)--(0,3.1) node[above,cadetgrey]{Good};
    \node[left,cadetgrey] at (0,0) {Easy};
    \node[below,cadetgrey] at (0,0) {Poor};
    \draw[->,thick,cadetgrey] (0,0)--(3,0) node[right]{Difficult};
    \draw[->,thick,cadetgrey] (0,0)--(0,3.1) node[above,cadetgrey]{Good};
    \draw[fill=gray] (2.56,1.5) circle [radius=0.05] node[left] {Python}; 
    \draw[fill=gray] (3,2) circle [radius=0.05] node[above] {\cpp};
    \end{tikzpicture}
    }
    \subfigure[Conjugate gradient]{
    \begin{tikzpicture}
    \draw[help lines, color=gray!30, dashed] (-0.1,-0.1) grid (2.9,2.9);
    \draw[->,thick,cadetgrey] (0,0)--(3,0) node[right]{Difficult};
    \draw[->,thick,cadetgrey] (0,0)--(0,3.1) node[above,cadetgrey]{Good};
    \draw[->,thick,cadetgrey] (0,0)--(3,0) node[right]{Difficult};
    \draw[->,thick,cadetgrey] (0,0)--(0,3.1) node[above,cadetgrey]{Good};
    \node[left,cadetgrey] at (0,0) {Easy};
    \node[below,cadetgrey] at (0,0) {Poor};
    \draw[fill=black] (2.608695652,3) circle [radius=0.05] node[left] {Python}; 
    \draw[fill=black] (3,3) circle [radius=0.05] node[above] {\cpp}; 
    \end{tikzpicture}
    }
    \subfigure[Heat equation solver]{
    \begin{tikzpicture}
    \draw[help lines, color=gray!30, dashed] (-0.1,-0.1) grid (2.9,2.9);
    \draw[->,thick,cadetgrey] (0,0)--(3,0) node[right]{Difficult};
    \draw[->,thick,cadetgrey] (0,0)--(0,3.1) node[above,cadetgrey]{Good};
    \draw[help lines, color=gray!30, dashed] (-0.1,-0.1) grid (2.9,2.9);
    \draw[->,thick,cadetgrey] (0,0)--(3,0) node[right]{Difficult};
    \draw[->,thick,cadetgrey] (0,0)--(0,3.1) node[above,cadetgrey]{Good};
    \node[left,cadetgrey] at (0,0) {Easy};
    \node[below,cadetgrey] at (0,0) {Poor};
    \draw[fill=azure] (2.7,2) circle [radius=0.05] node[left] {Python}; 
    \draw[fill=azure] (3,1.9) circle [radius=0.05] node[below] {\cpp};
    \end{tikzpicture}
    }
    \caption{On the x-axis, classification of the effort to translate Fortran code to Python and \cpp\ by a human developer from \textbf{easy} to \textbf{difficult} using Constructive Cost Model (COCOMO). On the y-axis, the quality of the translated code from \textbf{poor} to \textbf{good} was computed using the attributes: compilation, runtime, and correctness.}
    \label{fig:translation:quality:scatter}
\end{figure}

\subsection{Matlab to Python and R}
Table~\ref{tab:translation:results} (B) shows whether the code executed and provided the correct results. For the integration problem, both translated codes had runtime issues. However, after fixing issues, both codes produced the correct results. For the conjugate gradient problem, the Python code had runtime issues but produced correct results. The translated R coded worked. For the parallel heat equation solver problem, both Python and R codes had runtime issues. After fixing the runtime errors, only the Python code produced the correct results.

Figures~\ref{fig:translation:difference}(c-d) show the number of changes between the generated and translated codes. For the numerical integration problem, for the translation from Matlab to Python, the deleted lines were fewer than either the common and changed lines. This might be because Matlab and Python have similar programming language elements. For the translation from MatLab to R for the numerical integration, many lines were deleted or changed. For the conjugate gradient problem, for both Python and R, many lines were changed. For the parallel heat equation solver, only a few lines were changed or deleted for R while more than half were changed or deleted for Python.

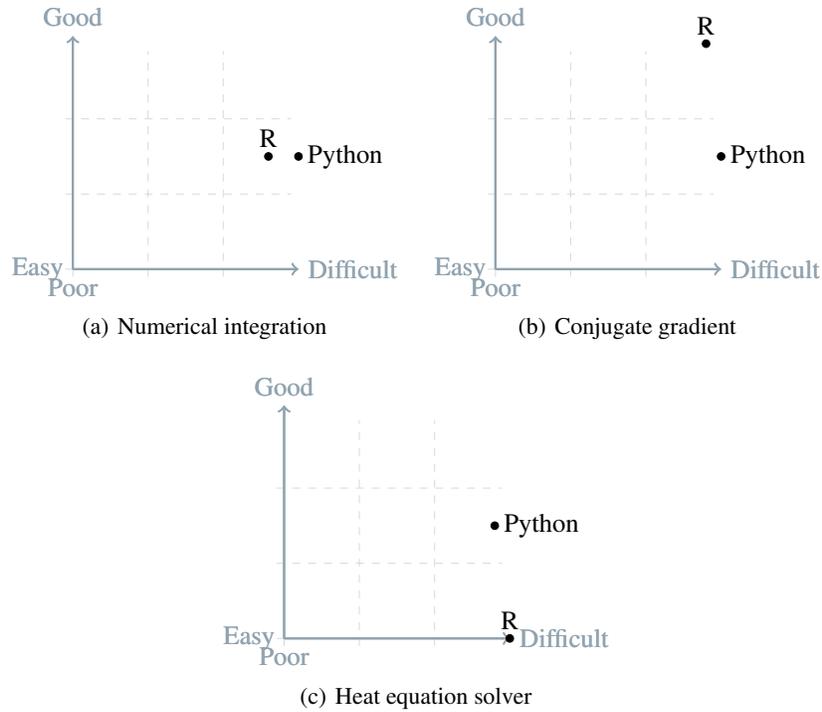
\begin{figure}[tb]
    \centering
    \subfigure[Numerical integration]{
    \begin{tikzpicture}
    \draw[help lines, color=gray!30, dashed] (-0.1,-0.1) grid (2.9,2.9);
    \draw[->,thick,cadetgrey] (0,0)--(3,0) node[right]{Difficult};
    \draw[->,thick,cadetgrey] (0,0)--(0,3.1) node[above,cadetgrey]{Good};
    \node[left,cadetgrey] at (0,0) {Easy};
    \node[below,cadetgrey] at (0,0) {Poor};
    \draw[->,thick,cadetgrey] (0,0)--(3,0) node[right]{Difficult};
    \draw[->,thick,cadetgrey] (0,0)--(0,3.1) node[above,cadetgrey]{Good};
    \node[left,cadetgrey] at (0,0) {Easy};
    \node[below,cadetgrey] at (0,0) {Poor};
    \draw[fill=black] (3,1.5) circle [radius=0.05] node[right] {Python}; 
    \draw[fill=black] (2.6,1.5) circle [radius=0.05] node[above] {R}; 
    \end{tikzpicture}
    }
    \subfigure[Conjugate gradient]{
    \begin{tikzpicture}
    \draw[help lines, color=gray!30, dashed] (-0.1,-0.1) grid (2.9,2.9);
    \draw[->,thick,cadetgrey] (0,0)--(3,0) node[right]{Difficult};
    \draw[->,thick,cadetgrey] (0,0)--(0,3.1) node[above,cadetgrey]{Good};
    \node[left,cadetgrey] at (0,0) {Easy};
    \node[below,cadetgrey] at (0,0) {Poor};
    \draw[->,thick,cadetgrey] (0,0)--(3,0) node[right]{Difficult};
    \draw[->,thick,cadetgrey] (0,0)--(0,3.1) node[above,cadetgrey]{Good};
    \node[left,cadetgrey] at (0,0) {Easy};
    \node[below,cadetgrey] at (0,0) {Poor};
    \draw[fill=black] (3,1.5) circle [radius=0.05] node[right] {Python}; 
    \draw[fill=black] (2.8,3) circle [radius=0.05] node[above] {R}; 
    \end{tikzpicture}
    }
    \subfigure[Heat equation solver]{
    \begin{tikzpicture}
    \draw[help lines, color=gray!30, dashed] (-0.1,-0.1) grid (2.9,2.9);
    \draw[->,thick,cadetgrey] (0,0)--(3,0) node[right]{Difficult};
    \draw[->,thick,cadetgrey] (0,0)--(0,3.1) node[above,cadetgrey]{Good};
    \node[left,cadetgrey] at (0,0) {Easy};
    \node[below,cadetgrey] at (0,0) {Poor};
    \draw[->,thick,cadetgrey] (0,0)--(3,0) node[right]{Difficult};
    \draw[->,thick,cadetgrey] (0,0)--(0,3.1) node[above,cadetgrey]{Good};
    \node[left,cadetgrey] at (0,0) {Easy};
    \node[below,cadetgrey] at (0,0) {Poor};
    \draw[fill=black] (2.8,1.5) circle [radius=0.05] node[right] {Python}; 
    \draw[fill=black] (3,0) circle [radius=0.05] node[above] {R}; 
    \end{tikzpicture}
    }
    \caption{On the x-axis, classification of the effort to translate Matlab to Python and R codes by a human developer from \textbf{easy} to \textbf{difficult} using the Constructive Cost Model (COCOMO). On the y-axis, the quality of the generated code from \textbf{poor} to \textbf{good} using the attributes: compilation, runtime, and correctness.}
    \label{fig:translation:quality:scatter:matlab}
\end{figure}

\section{Performance of the translated codes}
\label{sec:code:performance:translated}
For the translated codes, the performance of the conjugate gradient and the performance of the parallel heat equation solver were analyzed.  Figure~\ref{fig:translated:scaling} shows the results. For all runs, we plot the average run time among five runs. Figure~\ref{fig:translated:scaling:cg} shows the scaling for the conjugate gradient example for increasing matrix sizes. Figure~\ref{fig:translated:scaling:heat} shows the scaling for the parallel heat equation solver with increasing number of cores. For the \cpp\ code the scaled up to three cores. This behavior is similar to the codes generated with DeepSeek in the author's previous work~\cite{nader2025llmhpcbenchmarkingdeepseeks}. For Python, we can only show runs on a single core, since the model removed all parallelism while translating the code. All runs were done using an AMD\ EPYC\ 7742\ 64-Core Processor with 2 sockets and 128 cores in total.

\begin{figure}[tb]
    \centering
     \subfigure[\label{fig:translated:scaling:cg}]{
\resizebox{0.45\textwidth}{!}
        {
\begin{tikzpicture}
\begin{axis}[ylabel=average time (s),grid,legend,title=Conjugate gradient,ymode=log,log basis y={2},xmode=log,
       log basis x={2},legend pos=north west,xlabel=matrix size $(N \times N)$]
\addplot[black,mark=square*] table [x=matrix_size (NxN), y=avg(s), col sep=comma] {translation_cg_cpp.csv};
\addplot[black,mark=*] table [x=matrix_size (NxN), y=avg(s), col sep=comma] {translation_cg_python.csv};
\legend{\cpp,Python}
\end{axis}
\end{tikzpicture}
}
}
 \subfigure[\label{fig:translated:scaling:heat}]{
\resizebox{0.45\textwidth}{!}
        {
\begin{tikzpicture}
\begin{axis}[title=Parallel heat equation solver,xlabel=\# cores,ylabel=average time (s),grid,ymode=log,log basis y={2},xmode=log,log basis x={2},legend,legend pos=north west]
\addplot[black,mark=square*] table [x=omp_threads, y=avg(s), col sep=comma] {translation_omp_heat_shared_translated_cpp.csv};
\addplot[black,mark=*] table [x=omp_threads, y=avg(s), col sep=comma] {translation_omp_heat_shared_fixed_python.csv};
\legend{\cpp,Python}
\end{axis}
\end{tikzpicture}   
}
}
    \caption{Performance of the generated codes: \subref{fig:translated:scaling:cg} conjugate gradient with increasing matrix sizes on a single core and \subref{fig:translated:scaling:heat} parallel heat equation solver with increasing number of cores and a fixed problem size of 100,000 nodes and 10,000 time steps. For the Python code the problem size had to be reduced to 1024 nodes since the larger size took too long.}
    \label{fig:translated:scaling}
\end{figure}

\section{Conclusion and Outlook}
\label{sec:conclusion}

Based on our experiments, most of the current open source AI models cannot be trusted to write or translate even simple codes like the ones we have presented. Considerably more work, training, and enlarging the relevant models.

In the sections where accelerator code was studied, only the generation of CUDA code for NVIDIA GPUs was considered. However, AMD GPUs using HIP and Intel GPUs using SYCL are available. Since the generated CUDA code was not very promising, these two other GPU types were not studied. It may be that translating to kokkos (a performance portable framework) would be the best generic approach.


Alternatively, we could try providing the models with descriptions of the bugs and ask the model to fix the code. To avoid requiring human interaction, we could implement a fully automated workflow where an agent would find bugs and ask the model to fix them similar to what was proposed by Collini et al. in~\cite{collini2024c2hlsc}. \change{Another possible future direction is to investigate in more detail why the shared-memory code did not scale well.} This is conditioned by the quality of the code produced by AI models, which can be improved using supervised and unsupervised fine-tuning approaches. Recent studies have demonstrated significant improvements in both the correctness~\cite{shojaee2023execution} and performance~\cite{yang2024acecode} of generated codes. Building on these promising results, future research should investigate the potential of current fine-tuning methodologies for producing optimized HPC programs.

\section*{Supplementary materials}
All the generated material for this work available on GitHub\footnote{\url{https://github.com/diehlpkpapers/ai-journal-paper/}} or on Zenodo~\cite{}, respectively. The generated material using ChatGPT from the authors previous work~\cite{diehl2024evaluating} is available on GitHub\footnote{\url{https://github.com/diehlpkpapers/heat-ai}} or on Zenodo~\cite{noujoudnader_2025_15400128}, respectively. The generated material using DeepSeek from the authors previous work~\cite{nader2025llmhpcbenchmarkingdeepseeks} is available on GitHub\footnote{\url{https://github.com/NoujoudNader/AiCode_DeepSeek}} or on Zenodo~\cite{noujoudnader_2025_14968600}, respectively.

\section*{Acknowledgments}
This work was supported by the U.S. Department of Energy through the Los Alamos National Laboratory. Los Alamos National Laboratory is operated by Triad National Security, LLC, for the National Nuclear Security Administration of U.S. Department of Energy (Contract No. 89233218CNA000001). LA-UR-25-22576 (Revision 1)

\appendix

\section{Showcase of observed common errors}

\subsection{Undefined values and functions}

For the generated heat equation solver and for other codes, we obtained similar error messages. 
\begin{lstlisting}[language=bash]
python3 heat-shared.py
Traceback (most recent call last):
  File "/Users/diehlpk/git/ai-journal-paper/python/heat-shared.py", line 79, in <module>
    u = solve_heat_equation(u0, dx, dt, alpha, t_end, num_procs)
NameError: name 'dt' is not defined
\end{lstlisting}
The left listing shows the generated code and the right listing shows the fixed code defining the variable \lstinline[language=python]{dt}.\\
\hspace{0.25cm}
\begin{minipage}{0.45\textwidth}
\begin{lstlisting}[language=python]
# Problem parameters
    L = 1.0
    nx = 100
    dx = L / (nx - 1)
    t_end = 0.1
    alpha = 0.1
    num_procs = 4
\end{lstlisting}
\end{minipage}
\hfill
\begin{minipage}{0.45\textwidth}
\begin{lstlisting}[language=python]
# Problem parameters
    L = 1.0
    nx = 100
    dx = L / (nx - 1)
    t_end = 0.1
    alpha = 0.1
    num_procs = 4
    dt = 0.1 / 1000 
\end{lstlisting}
\end{minipage}\\
These issues were easy to address, since the variable or function just needed to be declared and initialized. However, knowledge about the physical model was required to choose a value for dt. We used the $0.1/1000$ since this was added by the model for some of the other codes. 

\subsection{Usage of incorrect constraints}
The AI model used incorrect constraints for some of the generated (and translated) codes. For the FORTRAN code, we observed the following error for the conjugate gradient method
\begin{lstlisting}[language=bash]
/Users/diehlpk/git/ai-journal-paper/fortran/cg.f90:93:8:

   93 |         b(j) = b(j) - factor * b(i)
      |        1
Error: Dummy argument 'b' with INTENT(IN) in variable definition context (assignment) at (1)
make[2]: *** [fortran/CMakeFiles/cg.dir/cg.f90.o] Error 1
make[1]: *** [fortran/CMakeFiles/cg.dir/all] Error 2
\end{lstlisting}
The left listing shows the generated code and the right listing shows the fixed code defining the variable \lstinline[language=python]{dt}.\\
\hspace{0.25cm}
\begin{minipage}{0.45\textwidth}
\begin{lstlisting}[language=fortran]
SUBROUTINE gaussian_elimination(A, b, x)
    REAL, DIMENSION(:,:), INTENT(IN) :: A
    REAL, DIMENSION(:), INTENT(IN) :: b
    ! Do some work 
    b(j) = b(j) - factor * b(i)
\end{lstlisting}
\end{minipage}
\hfill
\begin{minipage}{0.45\textwidth}
\begin{lstlisting}[language=fortran]
SUBROUTINE gaussian_elimination(A, b, x)
    REAL, DIMENSION(:,:), INTENT(IN) :: A
    REAL, DIMENSION(:), INTENT(OUT) :: b
    ! Do some work 
    b(j) = b(j) - factor * b(i)
\end{lstlisting}
\end{minipage}
The issue was that \lstinline[language=fortran]{INTENT(IN) :: b} was declared with the constraint \lstinline[language=fortran]{INTENT(IN)} which means that the variable is constant and will not be changed. However, later in the subroutine the variable is changed. The fix was to change the constraint to \lstinline[language=fortran]{INTENT(OUT)}.

\subsection{Multiple declarations}
For some of the codes, variables were declared multiple times with the same name. This resulted in the following error:
\begin{lstlisting}[language=bash]
[ 94%] Building CXX object translation/cpp/CMakeFiles/heat_shared_translated.dir/heat-shared.cpp.o
/Users/diehlpk/git/ai-journal-paper/translation/cpp/heat-shared.cpp:12:18: error: expected unqualified-id
   12 |     const double M_PI = 4 * atan(1);
      |                  ^
/Applications/Xcode.app/Contents/Developer/Platforms/MacOSX.platform/Developer/SDKs/MacOSX15.1.sdk/usr/include/math.h:710:21: note: expanded from macro 'M_PI'
  710 | #define M_PI        3.14159265358979323846264338327950288   /* pi             */
      |                     ^
1 error generated.
\end{lstlisting}
The left listing shows the generated code and the right listing shows the fixed code defining the variable \lstinline[language=python]{dt}.\\
\hspace{0.25cm}
\begin{minipage}{0.45\textwidth}
\begin{lstlisting}[language=c++]
#include <cmath>

const double M_PI = 4 * atan(1);
\end{lstlisting}
\end{minipage}
\hfill
\begin{minipage}{0.45\textwidth}
\begin{lstlisting}[language=c++]
#include <cmath>

//const double M_PI = 4 * atan(1);
\end{lstlisting}
\end{minipage}
The issue was that \lstinline[language=c++]{M_PI} is declared in the header \lstinline[language=c++]{#include <cmath>} and was later declared again. The fix was to comment the second declaration out and use the one declared in the header.

\subsection{Choosing physical values }
For some generated codes, the model chose incorrect values for some of the physical constants. An example was that the time step size \lstinline[language=c++]{dt} was set to too large of value and the simulation became unstable. To address the correctness, the time step size had to be changed to a lower value. Here, some understanding of the numerics was needed to address this issue. \\
The left listing shows the generated code and the right listing shows the fixed code defining the variable \lstinline[language=python]{dt}.\\
\hspace{0.25cm}
\begin{minipage}{0.45\textwidth}
\begin{lstlisting}[language=c++]
double dt = 0.1;
\end{lstlisting}
\end{minipage}
\hfill
\begin{minipage}{0.45\textwidth}
\begin{lstlisting}[language=c++]
double dt = 0.0001;
\end{lstlisting}
\end{minipage}









\bibliographystyle{Bibliography_Style}

\bibliography{References}
\end{document}